\documentclass[12pt,twoside]{article}
\usepackage{amssymb,amsmath,mathrsfs,deluxetable,hogg_endnotes,natbib}
\usepackage{float,graphicx}

\newcommand{\numberparagraphs}{}

\makeatletter
\newsavebox{\tempbox}
\newcommand{\@makefigcaption}[2]{%
\vspace{10pt}{#1.--- #2\par}}%
\renewcommand{\figure}{\let\@makecaption\@makefigcaption\@float{figure}}
\makeatother

\newcommand{\exampleplot}[1]{%
\begin{center}%
\includegraphics[width=0.5\textwidth]{#1}%
\end{center}%
}
\newcommand{\exampleplottwo}[2]{%
\begin{center}%
\includegraphics[width=0.5\textwidth]{#1}%
\includegraphics[width=0.5\textwidth]{#2}%
\end{center}%
}

\newcommand{\notenglish}[1]{\textsl{#1}}
\newcommand{\aposteriori}{\notenglish{a~posteriori}}
\newcommand{\apriori}{\notenglish{a~priori}}
\newcommand{\adhoc}{\notenglish{ad~hoc}}
\newcommand{\etal}{\notenglish{et al.}}

\newcommand{\documentname}{document}
\newcommand{\sectionname}{Section}
\newcommand{\equationname}{equation}
\newcommand{\figurenames}{\figurename s}
\newcommand{\problemname}{Exercise}
\newcommand{\problemnames}{\problemname s}

\newcommand{\notename}{note}

\newcommand{\note}[1]{\endnote{#1}}
\def\enotesize{\normalsize}

\newcounter{problem}
\newenvironment{problem}{\paragraph{\problemname~\theproblem:}\refstepcounter{problem}}{}
\newcommand{\affil}[1]{{\footnotesize\textsl{#1}}}

\newcommand{\mmatrix}[1]{\boldsymbol{#1}}
\newcommand{\inverse}[1]{{#1}^{-1}}
\newcommand{\transpose}[1]{{#1}^{\scriptscriptstyle \top}}
\newcommand{\mA}{\mmatrix{A}}
\newcommand{\mAT}{\transpose{\mA}}
\newcommand{\mC}{\mmatrix{C}}
\newcommand{\mCinv}{\inverse{\mC}}
\newcommand{\mQ}{\mmatrix{Q}}
\newcommand{\mS}{\mmatrix{S}}
\newcommand{\mX}{\mmatrix{X}}
\newcommand{\mY}{\mmatrix{Y}}

\newcommand{\mZ}{\mmatrix{Z}}
\newcommand{\vhat}{\mmatrix{\hat{v}}}

\newcommand{\parametervector}[1]{\mmatrix{#1}}
\newcommand{\pvtheta}{\parametervector{\theta}}

\newcommand{\setofall}[3]{\{{#1}\}_{{#2}}^{{#3}}}
\newcommand{\allq}{\setofall{q_i}{i=1}{N}}
\newcommand{\allx}{\setofall{x_i}{i=1}{N}}
\newcommand{\ally}{\setofall{y_i}{i=1}{N}}
\newcommand{\allxy}{\setofall{x_i,y_i}{i=1}{N}}
\newcommand{\allsigmay}{\setofall{\sigma_{yi}^2}{i=1}{N}}
\newcommand{\allS}{\setofall{\mS_i}{i=1}{N}}

\renewcommand{\d}{\mathrm{d}}
\newcommand{\like}{\mathscr{L}}
\newcommand{\pfg}{p_{\mathrm{fg}}}
\newcommand{\pbg}{p_{\mathrm{bg}}}
\newcommand{\Pbad}{P_{\mathrm{b}}}
\newcommand{\Ybad}{Y_{\mathrm{b}}}
\newcommand{\Vbad}{V_{\mathrm{b}}}
\newcommand{\bperp}{b_{\perp}}
\newcommand{\mean}[1]{\left<{#1}\right>}
\newcommand{\meanZ}{\mean{\mZ}}

\renewcommand{\MakeUppercase}[1]{#1}
\begin{document}
\thispagestyle{plain}\raggedbottom
\section*{Data analysis recipes:\ \\
  Fitting a model to data\footnotemark}

\footnotetext{%
  The \notename s begin on page~\pageref{note:first}, including the
  license\note{\label{note:first}%
    Copyyright 2010 by the authors.  You may copy and distribute this
    document provided that you make no changes to it whatsoever.}  and
  the acknowledgements\note{%
    Above all we owe a debt to Sam Roweis (Toronto \& NYU, now
    deceased), who taught us to find and optimize justified scalar
    objective functions.  He will be sorely missed by all three of us. 
    In addition it is a pleasure to thank Mike
    Blanton (NYU), Scott Burles (D. E. Shaw), Daniel Foreman-Mackey
    (Queens), Brandon C. Kelly (Harvard), Iain Murray (Edinburgh),
    Hans-Walter Rix (MPIA), and David Schiminovich (Columbia) for
    valuable comments and discussions.  This research was partially
    supported by NASA (ADP grant NNX08AJ48G), NSF (grant AST-0908357),
    and a Research Fellowship of the Alexander von Humboldt
    Foundation.  This research made use of the Python programming
    language and the open-source Python packages scipy, numpy, and
    matplotlib.}.}

\noindent
David~W.~Hogg\\
\affil{Center~for~Cosmology~and~Particle~Physics, Department~of~Physics, New York University}\\
\affil{Max-Planck-Institut f\"ur Astronomie, Heidelberg}
\\[1ex]
Jo~Bovy\\
\affil{Center~for~Cosmology~and~Particle~Physics, Department~of~Physics, New York University}
\\[1ex]
Dustin~Lang\\
\affil{Department of Computer Science, University of Toronto}\\
\affil{Princeton University Observatory}

\begin{abstract}
  We go through the many considerations involved in fitting a model to
  data, using as an example the fit of a straight line to a set of
  points in a two-dimensional plane.  Standard weighted least-squares
  fitting is only appropriate when there is a dimension along which
  the data points have negligible uncertainties, and another along
  which all the uncertainties can be described by Gaussians of known
  variance; these conditions are rarely met in practice.  We consider
  cases of general, heterogeneous, and arbitrarily covariant
  two-dimensional uncertainties, and situations in which there are bad
  data (large outliers), unknown uncertainties, and unknown but
  expected intrinsic scatter in the linear relationship being fit.
  Above all we emphasize the importance of having a ``generative
  model'' for the data, even an approximate one.  Once there is a
  generative model, the subsequent fitting is non-arbitrary because
  the model permits direct computation of the likelihood of the
  parameters or the posterior probability distribution.  Construction
  of a posterior probability distribution is indispensible if there
  are ``nuisance parameters'' to marginalize away.
\end{abstract}

\numberparagraphs

It is conventional to begin any scientific document with an
introduction that explains why the subject matter is important.  Let
us break with tradition and observe that in almost all cases in which
scientists fit a straight line to their data, they are doing something
that is simultaneously \emph{wrong} and \emph{unnecessary}.  It is
wrong because circumstances in which a set of two dimensional
measurements---outputs from an observation, experiment, or
calculation---are truly drawn from a narrow, linear relationship is
exceedingly rare.  Indeed, almost any transformation of coordinates
renders a truly linear relationship non-linear.  Furthermore, even if
a relationship \emph{looks} linear, unless there is a confidently held
theoretical reason to believe that the data are generated from a
linear relationship, it probably isn't linear in detail; in these
cases fitting with a linear model can introduce substantial systematic
error, or generate apparent inconsistencies among experiments that are
intrinsically consistent.

Even if the investigator doesn't care that the fit is wrong, it is
likely to be unnecessary.  Why?  Because it is rare that, given a
complicated observation, experiment, or calculation, the important
\emph{result} of that work to be communicated forward in the
literature and seminars is the \emph{slope and intercept} of a
best-fit line!  Usually the full distribution of data is much more
rich, informative, and important than any simple metrics made by
fitting an overly simple model.

That said, it must be admitted that one of the most effective ways to
communicate scientific results is with catchy punchlines and compact,
approximate representations, even when those are unjustified and
unnecessary.  It can also sometimes be useful to fit a simple model to
predict new data, given existing data, even in the absence of a
physical justification for the fit.\note{When fitting is used for
  data-driven prediction---that is, using existing data to predict the
  properties of new data not yet acquired---the conditions for model
  applicability are weaker.  For example, a line fit by the standard
  method outlined in \sectionname~\ref{sec:standard} can be, under a
  range of assumptions, the best linear predictor of new data, even
  when it is not a good or appropriate model.  A full discussion of
  this, which involves model selection and clarity about exactly what
  is being predicted (it is different to predict $y$ given $x$ than to
  predict $x$ given $y$ or $x$ and $y$ together), is beyond the scope
  of this \documentname, especially since in astrophysics and other
  observational sciences it is relatively rare that data prediction is
  the goal of fitting.}  For these reasons---and the reason that in
rare cases the fit \emph{is} justifiable and essential---the problem
of fitting a line to data comes up very frequently in the life of a
scientist.

It is a miracle with which we hope everyone reading this is familiar
that \emph{if} you have a set of two-dimensional points $(x,y)$ that
depart from a perfect, narrow, straight line $y=m\,x+b$ only by the
addition of Gaussian-distributed noise of known amplitudes in the $y$
direction only, \emph{then} the maximum-likelihood or best-fit line
for the points has a slope $m$ and intercept $b$ that can be obtained
justifiably by a perfectly linear matrix-algebra operation known as
``weighted linear least-square fitting''.  This miracle deserves
contemplation.

Once any of the input assumptions is violated (and note there are many
input assumptions), all bets are off, and there are no consensus
methods used by scientists across disciplines.  Indeed, there is a
large literature that implies that the violation of the assumptions
opens up to the investigator a wide range of possible \emph{options}
with little but aesthetics to distinguish them!  Most of these methods
of ``linear regression'' (a term we avoid) performed in science are
either unjustifiable or else unjustified in almost
all situations in which they are used.\note{The term ``linear
  regression'' is used in many different contexts, but in the most
  common, they seem to mean ``linear fitting without any use of the
  uncertainties''.  In other words, linear regression is usually just
  the procedure outlined in this \sectionname, but with the covariance
  matrix $\mC$ set equal to the identity matrix.  This is equivalent
  to fitting under the assumption not just that the $x$-direction
  uncertainties are negligible, but also that all of the $y$-direction
  uncertainties are simultaneously identical and Gaussian.

  Sometimes the term ``linear regression'' is meant to cover a much
  larger range of options; in our field of astrophysics there have
  been attempts to test them all (for example, see \citealt{isobe90}).
  This kind of work---testing \adhoc\ procedures on simulated
  data---can only be performed if you are willing to simultaneously
  commit (at least) two errors: \emph{First,} the linear regression
  methods are \emph{procedures}; they do not (necessarily) optimize
  anything that a scientist would care about. That is, the vast
  majority of procedures fail to produce a best-fit line in any
  possible sense of the word ``best''.  Of course, some procedures do
  produce a best-fit line under some assumptions, but scientific
  procedures should flow from assumptions and not the other way
  around.  Scientists do not trade in procedures, they trade in
  objectives, and choose procedures only when they are demonstrated to
  optimize their objectives (we very much hope).  \emph{Second,} in
  this kind of testing, the investigator must decide which procedure
  ``performs'' best by applying the procedures to sets of
  \emph{simulated data}, where truth---the true generative model---is
  known.  All that possibly can be shown is which procedure does well
  at reproducing the parameters of the model with which the simulated
  data are generated!  This has no necessary relationship with
  anything by which the \emph{real} data are generated.  Even if the
  generative model for the artificial data matches perfectly the
  generation of the true data, it is still much better to analyze the
  likelihood created by this generative model than to explore
  procedure space for a procedure that comes close to doing that by
  accident---unless, perhaps, when computational limitations demand
  that only a small number of simple operations be performed on the
  data.\label{note:regression}} Perhaps because of this plethora of
options, or perhaps because there is no agreed-upon bible or cookbook
to turn to, or perhaps because most investigators would rather make
some stuff up that works ``good enough'' under deadline, or perhaps
because many realize, deep down, that much fitting is really
unnecessary anyway, there are some egregious procedures and associated
errors and absurdities in the literature.\note{Along the lines of
  ``egregious'', in addition to the complaints in the previous
  \notename, the above-cited paper \citep{isobe90} makes another
  substantial error: When deciding whether to fit for $y$ as a
  function of $x$ or $x$ as a function of $y$, the authors claim that
  the decision should be based on the ``physics'' of $x$ and $y$!
  This is the ``independent variable'' fallacy---the investigator
  thinks that $y$ is the independent variable because physically it
  ``seems'' independent (as in ``it is really the velocity that
  depends on the distance, not the distance that depends on the
  velocity'' and other statistically meaningless intuitions).  The
  truth is that this decision must be based on the \emph{uncertainty
    properties} of $x$ and $y$. If $x$ has much smaller uncertainties,
  then you must fit $y$ as a function of $x$; if the other way then
  the other way, and if neither has much smaller uncertainties, then
  that kind of linear fitting is invalid.  We have more to say about
  this generic situation in later \sectionname s.  What is written in
  \cite{isobe90}---by professional statisticians, we could add---is
  simply wrong.\label{note:fallacy}} When an investigator cites,
relies upon, or transmits a best-fit slope or intercept reported in
the literature, he or she may be propagating more noise than signal.

It is one of the objectives of this \documentname\ to promote an
understanding that if the data can be \emph{modeled} statistically
there is little arbitrariness.  It is another objective to promote
consensus; when the choice of method is not arbitrary, consensus will
emerge automatically.  We present below simple, straightforward,
comprehensible, and---above all---\emph{justifiable} methods for
fitting a straight line to data with general, non-trivial, and
uncertain properties.

This \documentname\ is part polemic (though we have tried to keep most
of the polemic in the \notename s), part attempt at establishing
standards, and part crib sheet for our own future re-use.  We
apologize in advance for some astrophysics bias and failure to review
basic ideas of linear algebra, function optimization, and practical
statistics.  Very good reviews of the latter exist in
abundance.\note{For getting started with modern data modeling and
  inference, the texts \cite{mackay}, \cite{press}, and \cite{sivia}
  are all very useful.}  We also apologize for not reviewing the
literature; this is a cheat-sheet not a survey.  Our focus is on the
specific problems of linear fitting that we so often face; the general
ideas appear in the context of these concrete and relatively realistic
example problems.  The reader is encouraged to do the \problemnames;
many of these ask you to produce plots which can be compared with the
\figurenames.

\section{Standard practice}\label{sec:standard}

You have a set of $N>2$ points $(x_i,y_i)$, with known Gaussian
uncertainties $\sigma_{yi}$ in the $y$ direction, and no uncertainty
at all (that is, perfect knowledge) in the $x$ direction.  You want to
find the function $f(x)$ of the form
\begin{equation}\label{eq:fofx}
f(x) = m\,x + b \quad ,
\end{equation}
where $m$ is the slope and $b$ is the intercept, that \emph{best fits}
the points.  What is meant by the term ``best fit'' is, of course,
very important; we are making a choice here to which we will return.
For now, we describe standard practice:

Construct the matrices
\begin{align}\label{eq:mY}
\mY &= \left[\begin{array}{c}
y_1 \\
y_2 \\
\cdots \\
y_N
\end{array}\right] \quad ,\\
\mA &= \left[\begin{array}{cc}
1 & x_1 \\
1 & x_2 \\
\multicolumn{2}{c}{\cdots} \\
1 & x_N
\end{array}\right] \quad ,\\
\mC & = \left[\begin{array}{cccc}
\sigma_{y1}^{2} & 0 & \cdots & 0 \\
0 & \sigma_{y2}^{2} & \cdots & 0 \\
\multicolumn{4}{c}{\cdots} \\
0 & 0 & \cdots & \sigma_{yN}^{2}
\end{array}\right] \quad ,\label{eq:covar}
\end{align}
where one might call $\mY$ a ``vector'', and the covariance matrix
$\mC$ could be generalized to the non-diagonal case in which there are
covariances among the uncertainties of the different points.  The
best-fit values for the parameters $m$ and $b$ are just the components
of a column vector $\mX$ found by
\begin{equation}\label{eq:lsf}
\left[\begin{array}{c} $b$ \\ $m$ \end{array}\right]
 = \mX = \inverse{\left[\mAT\,\mCinv\,\mA\right]}
  \,\left[\mAT\,\mCinv\,\mY\right] \quad .
\end{equation}
This is actually the simplest thing that can be written down that is
linear, obeys matrix multiplication rules, and has the right relative
sensitivity to data of different statistical significance.  It is not
just the simplest thing; it is the correct thing, when the assumptions
hold.\note{Even in situations in which linear fitting is appropriate,
  it is \emph{still} often done wrong!  It is not unusual to see the
  individual data-point uncertainty estimates ignored, even when they
  are known at least approximately.  It is also common for the problem
  to get ``transposed'' such that the coordinates for which
  uncertainties are negligible are put into the $\mY$ vector and the
  coordinates for which uncertainties are \emph{not} negligible are
  put into the $\mA$ matrix.  In this latter case, the procedure makes
  no sense at all really; it happens when the investigator thinks of
  some quantity ``really being'' the dependent variable, despite the
  fact that it has the smaller uncertainty.  For example there are
  mistakes in the Hubble expansion literature, because many have an
  intuition that the ``velocity depends on the distance'' when in fact
  velocities are measured better, so from the point of view of
  fitting, it is better to think of the distance depending on the
  velocity.  In the context of fitting, there is no meaning to the
  (ill-chosen) terms ``independent variable'' and ``dependent
  variable'' beyond the uncertainty or noise properties of the data.
  This is the independent variable fallacy mentioned in
  \notename~\ref{note:fallacy}.  In performing this standard fit, the
  investigator is effectively assuming that the $x$ values have
  negligible uncertainties; if they do \emph{not}, then the
  investigator is \emph{making a mistake}.}  It can be justified in
one of several ways; the linear algebra justification starts by noting
that you want to solve the equation
\begin{equation}
\mY = \mA\,\mX \quad ,
\end{equation}
but you can't because that equation is over-constrained.  So you
weight everything with the inverse of the covariance matrix (as you
would if you were doing, say, a weighted average), and then
left-multiply everything by $\mAT$ to reduce the dimensionality, and
then \equationname~(\ref{eq:lsf}) is the solution of that
reduced-dimensionality equation.

This procedure is not arbitrary; it minimizes an objective
function\note{As we mention in \notename~\ref{note:regression},
  scientific procedures should achieve justified objectives; this
  objective function makes that statement well posed.  The objective
  function is also sometimes called the ``loss function'' in the
  statistics literature.} $\chi^2$ (``chi-squared''), which is the
total squared error, scaled by the uncertainties, or
\begin{equation}\label{eq:chisquared}
\chi^2
 = \sum_{i=1}^N \frac{\left[y_i - f(x_i)\right]^2}{\sigma_{yi}^2}
 \equiv \transpose{\left[\mY-\mA\,\mX\right]}
 \,\mCinv\,\left[\mY-\mA\,\mX\right]
 \quad ,
\end{equation}
that is, \equationname~(\ref{eq:lsf}) yields the values for $m$ and
$b$ that minimize $\chi^2$.  Conceptually, $\chi^2$ is like a metric
distance in the data space.\note{The inverse covariance matrix
  appears in the construction of the $\chi^2$ objective function like
  a linear ``metric'' for the data space: It is used to turn a
  $N$-dimensional vector displacement into a scalar squared distance
  between the observed values and the values predicted by the
  model. This distance can then be minimized.  This idea---that the
  covariance matrix is a metric---is sometimes useful for thinking
  about statistics as a physicist; it will return in
  \sectionname~\ref{sec:twod} when we think about data for which there
  are uncertainties in both the $x$ and $y$ directions.

  Because of the purely quadratic nature of this metric distance, the
  standard problem solved in this \sectionname\ has a linear solution.
  Not only that, but the objective function ``landscape'' is also
  \emph{convex} so the optimum is global and the solution is unique.
  This is remarkable, and almost no perturbation of this problem (as
  we will see in later \sectionname s) has either of these properties,
  let alone both.  For this standard problem, the linearity and
  convexity permit a one-step solution with no significant engineering
  challenges, even when $N$ gets exceedingly large.  As we modify (and
  make more realistic) this problem, linearity and convexity will
  \emph{both} be broken; some thought will have to be put into the
  optimization.}  This, of course, is only one possible meaning of the
phrase ``best fit''; that issue is addressed more below.

When the uncertainties are Gaussian and their variances $\sigma_{yi}$
are correctly estimated, the matrix
$\inverse{\left[\mAT\,\mCinv\,\mA\right]}$ that appears in
\equationname~(\ref{eq:lsf}) is just the covariance matrix (Gaussian
uncertainty---uncertainty not error\note{There is a difference between
  ``uncertainty'' and ``error'': For some reason, the uncertainty
  estimates assigned to data points are often called ``errors'' or
  ``error bars'' or ``standard errors''.  That terminology is wrong:
  ``They are uncertainties, not errors; if they were \emph{errors}, we
  would have corrected them!''  DWH's mentor Gerry Neugebauer liked to
  say this; Neugebauer credits the quotation to physicist Matthew
  Sands.  Data come with \emph{uncertainties}, which are limitations
  of knowledge, not \emph{mistakes}.\label{note:error}}---variances on
the diagonal, covariances off the diagonal) for the parameters in
$\mX$.  We will have more to say about this in 
\sectionname~\ref{sec:uncertainty}.

\begin{deluxetable}{rrrrrr@{.}l}
\tablecolumns{7}
\tablehead{ID &$x$ & $y$ & $\sigma_y$ & $\sigma_x$ &  \multicolumn{2}{c}{$\rho_{xy}$}}
\tablewidth{0pt}
\startdata
1 & 201 & 592 & 61 & 9 & -0 & 84\\
2 & 244 & 401 & 25 & 4 & 0 & 31\\
3 & 47 & 583 & 38 & 11 & 0 & 64\\
4 & 287 & 402 & 15 & 7 & -0 & 27\\
5 & 203 & 495 & 21 & 5 & -0 & 33\\
6 & 58 & 173 & 15 & 9 & 0 & 67\\
7 & 210 & 479 & 27 & 4 & -0 & 02\\
8 & 202 & 504 & 14 & 4 & -0 & 05\\
9 & 198 & 510 & 30 & 11 & -0 & 84\\
10 & 158 & 416 & 16 & 7 & -0 & 69\\
11 & 165 & 393 & 14 & 5 & 0 & 30\\
12 & 201 & 442 & 25 & 5 & -0 & 46\\
13 & 157 & 317 & 52 & 5 & -0 & 03\\
14 & 131 & 311 & 16 & 6 & 0 & 50\\
15 & 166 & 400 & 34 & 6 & 0 & 73\\
16 & 160 & 337 & 31 & 5 & -0 & 52\\
17 & 186 & 423 & 42 & 9 & 0 & 90\\
18 & 125 & 334 & 26 & 8 & 0 & 40\\
19 & 218 & 533 & 16 & 6 & -0 & 78\\
20 & 146 & 344 & 22 & 5 & -0 & 56\\
\tablecomments{The full uncertainty covariance matrix for each data point is given by\\ $\left[\begin{array}{cc} \sigma_x^2 & \rho_{xy}\sigma_x\sigma_y\\\rho_{xy}\sigma_x\sigma_y & \sigma_y^2\end{array}\right]$.}
\label{table:data_allerr}
\enddata
\end{deluxetable}

\begin{figure}[!htbp]
\exampleplot{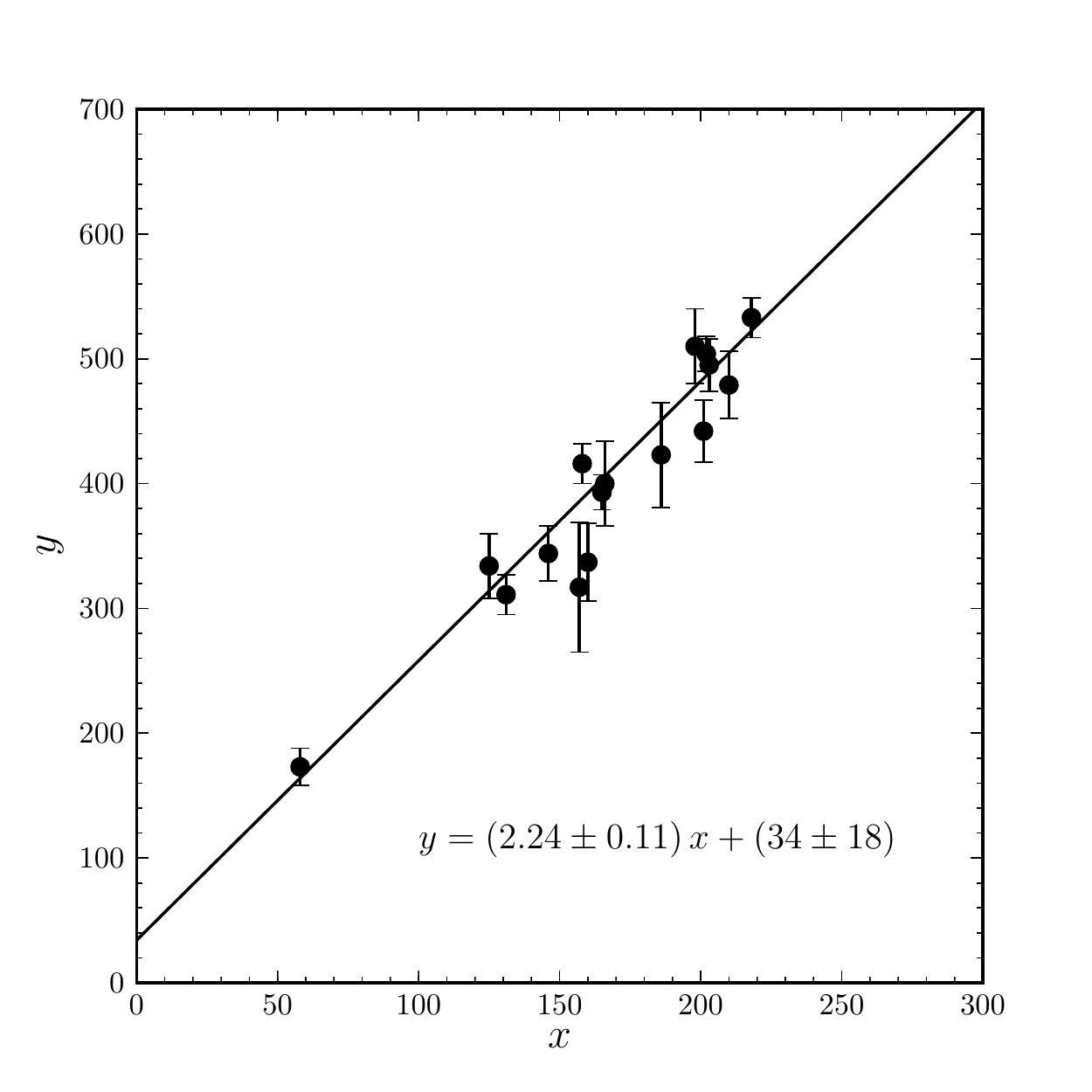}
\caption{Partial solution to \problemname~\ref{prob:easy}: The
standard weighted least-square fit.}\label{fig:easy}
\end{figure}

\begin{problem}\label{prob:easy}
Using the standard linear algebra method of this \sectionname, fit the
straight line $y=m\,x+b$ to the $x$, $y$, and $\sigma_y$ values for
data points 5 through 20 in \tablename~\ref{table:data_allerr} on
page~\pageref{table:data_allerr}.  That is, ignore the first four data
points, and also ignore the columns for $\sigma_x$ and $\rho_{xy}$.
Make a plot showing the points, their uncertainties, and the best-fit
line.  Your plot should end up looking like
\figurename~\ref{fig:easy}.  What is the standard uncertainty variance
$\sigma_m^2$ on the slope of the line?
\end{problem}

\begin{figure}[htbp!]
\exampleplot{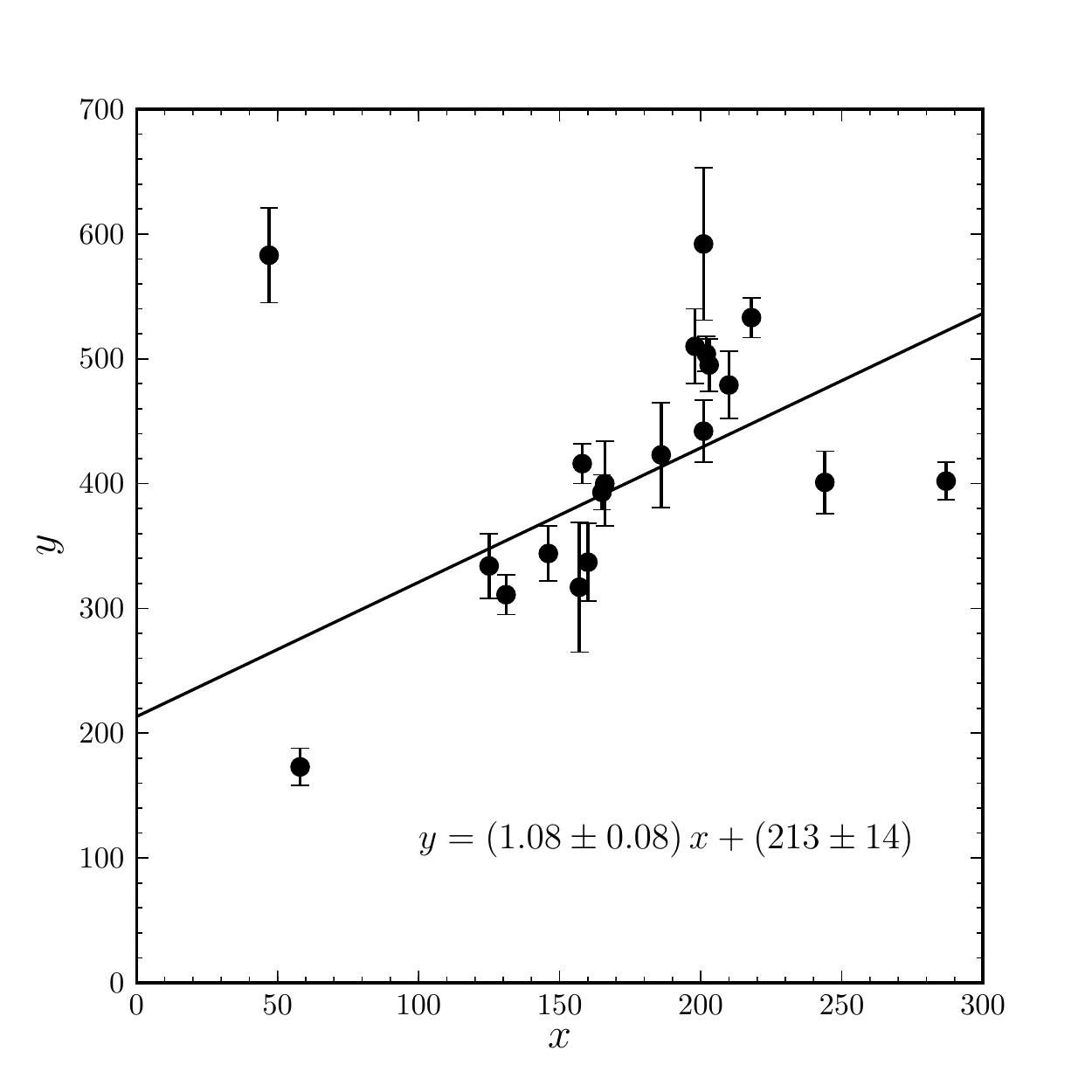}
\caption{Partial solution to \problemname~\ref{prob:standard}: The
standard weighted least-square fit but now on a larger data
set.}\label{fig:standard}
\end{figure}

\begin{problem}\label{prob:standard}
Repeat \problemname~\ref{prob:easy} but for all the data points in
\tablename~\ref{table:data_allerr} on
page~\pageref{table:data_allerr}.  Your plot should end up looking
like \figurename~\ref{fig:standard}.  What is the standard uncertainty
variance $\sigma_m^2$ on the slope of the line?  Is there anything you
don't like about the result?  Is there anything different about the
new points you have included beyond those used in
\problemname~\ref{prob:easy}?
\end{problem}

\begin{figure}[htbp!]
\exampleplot{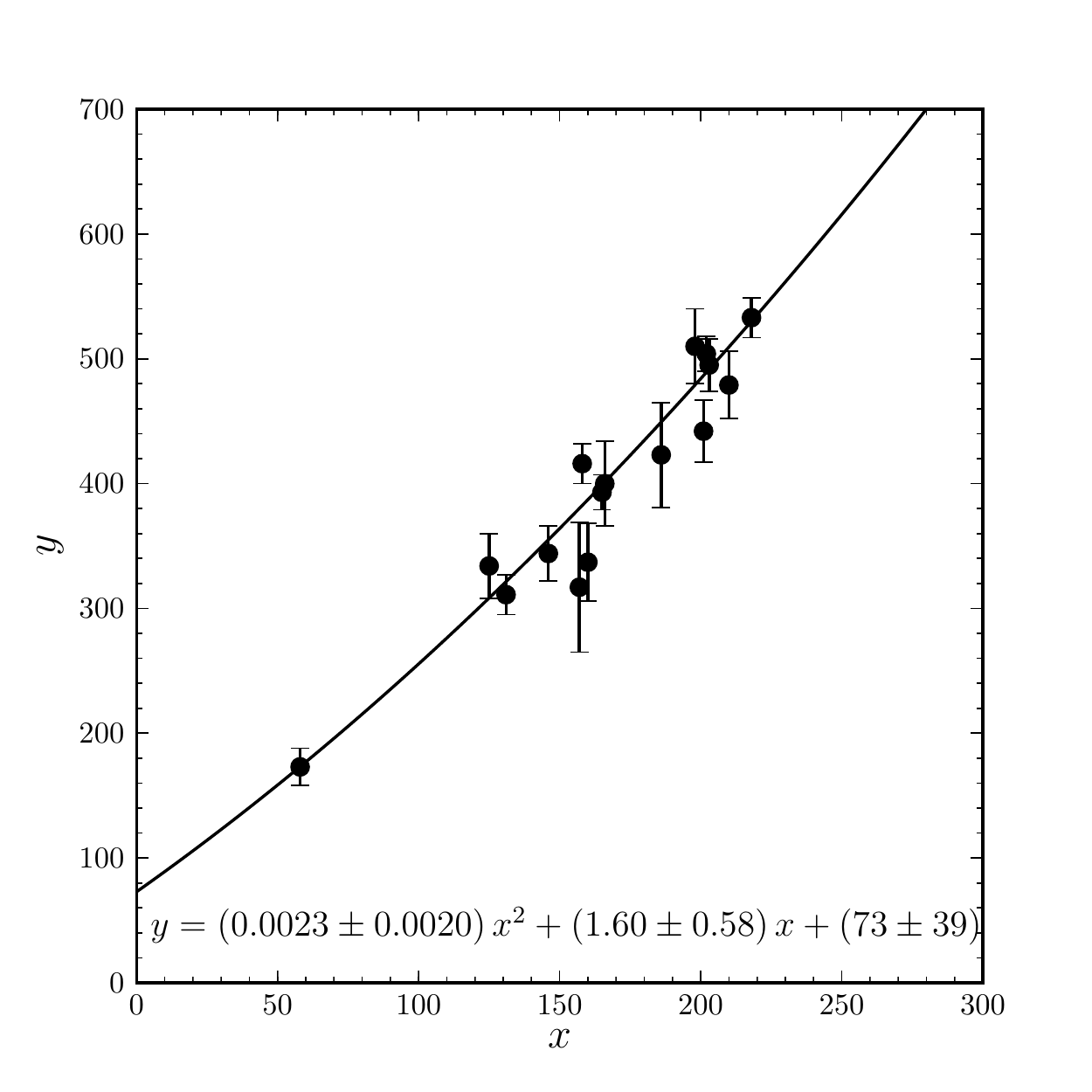}
\caption{Partial solution to \problemname~\ref{prob:quadratic}: The
standard weighted least-square fit but now with a second-order
(quadratic) model.}\label{fig:quadratic}
\end{figure}

\begin{problem}\label{prob:quadratic}
Generalize the method of this \sectionname\ to fit a general quadratic
(second order) relationship.  Add another column to matrix $\mA$
containing the values $x_i^2$, and another element to vector $\mX$
(call it $q$).  Then re-do \problemname~\ref{prob:easy} but
fitting for and plotting the best quadratic relationship
\begin{equation}
g(x) = q\,x^2 + m\,x + b \quad.
\end{equation}
Your plot should end up looking like \figurename~\ref{fig:quadratic}.
\end{problem}

\numberparagraphs

\section{The objective function}\label{sec:objective}

A scientist's justification of \equationname~(\ref{eq:lsf}) cannot
appeal purely to abstract ideas of linear algebra, but must originate
from the scientific question at hand.  Here and in what follows, we
will advocate that the only reliable procedure is to use all one's
knowledge about the problem to construct a (preferably) justified,
(necessarily) scalar (or, at a minimum, one-dimensional),
\emph{objective function} that represents monotonically the quality of
the fit.\note{In geometry or physics, a ``scalar'' is not just a
  quantity with a magnitude; it is a quantity with a magnitude that
  has certain transformation properties: It is invariant under some
  relevant transformations of the coordinates.  The objective ought to
  be something like a scalar in this sense, because we want the
  inference to be as insensitive to coordinate choices as possible.
  The standard $\chi^2$ objective function is a scalar in the sense
  that it is invariant to translations or rescalings of the data in
  the $\mY$ vector space defined in \equationname~(\ref{eq:mY}).  Of
  course a \emph{rotation} of the data in the $N$-dimensional space of
  the $\ally$ would be pretty strange, and would make the covariance
  matrix $\mC$ much more complicated, so perhaps the fact that the
  objective is literally a linear algebra scalar---this is clear from
  its matrix representation in
  \equationname~(\ref{eq:chisquared})---is something of a red
  herring.

  The more important thing about the objective is that it must have
  \emph{one magnitude only}.  Often when one asks scientists what they
  are trying to optimize, they say things like ``I am trying to
  maximize the resolution, minimize the noise, and minimize the error
  in measured fluxes''.  Well, those are good objectives, but
  \emph{you can't optimize them all}.  You either have to pick one, or
  make a convex combination of them (such as by adding them together
  in quadrature with weights).  This point is so obvious, it is not
  clear what else to say, except that many scientific investigations
  are so ill-posed that the investigator does not know what is being
  optimized.  When there is a generative model for the data, this
  problem doesn't arise.}  In this framework, fitting anything to
anything involves a scientific question about the objective function
representing ``goodness of fit'' and then a separate and subsequent
engineering question about how to \emph{find the optimum} and,
possibly, the posterior probability distribution function around that
optimum.\note{Of course, whether you are a frequentist or a Bayesian,
  the most scientifically responsible thing to do is not just
  \emph{optimize} the objective, but pass forward a description of the
  dependence of the objective on the parameters, so that other
  investigators can combine your results with theirs in a responsible
  way.  In the fully Gaussian case described in this \sectionname,
  this means passing forward the optimum \emph{and} second derivatives
  around that optimum (since a Gaussian is fully specified by its
  mode---or mean--and its second derivative at the mode).  In what
  follows, when things aren't purely Gaussian, more sophisticated
  outputs will be necessary.}  Note that in the previous
\sectionname\ we did \emph{not} proceed according to these rules;
indeed the procedure was introduced prior to the objective function,
and the objective function was not justified.

In principle, there are many choices for objective function.  But the
only procedure that is truly justified---in the sense that it leads to
interpretable probabilistic inference, is to make a \emph{generative
  model} for the data.  A generative model is a parameterized,
quantitative description of a statistical procedure that could
reasonably have generated the data set you have.  In the case of the
straight line fit in the presence of known, Gaussian uncertainties in
one dimension, one can create this generative model as follows:
Imagine that the data \emph{really do} come from a line of the form $y
= f(x) = m\,x+b$, and that the only reason that any data point
deviates from this perfect, narrow, straight line is that to each of
the true $y$ values a small $y$-direction offset has been added, where
that offset was drawn from a Gaussian distribution of zero mean and
known variance $\sigma_y^2$.  In this model, given an independent
position $x_i$, an uncertainty $\sigma_{yi}$, a slope $m$, and an
intercept $b$, the frequency distribution $p(y_i|x_i,\sigma_{yi},m,b)$
for $y_i$ is
\begin{equation}\label{eq:objectivei}
p(y_i|x_i,\sigma_{yi},m,b) = \frac{1}{\sqrt{2\,\pi\,\sigma_{yi}^2}}
 \,\exp\left(-\frac{[y_i - m\,x_i - b]^2}{2\,\sigma_{yi}^2}\right) \quad ,
\end{equation}
where this gives the expected frequency (in a hypothetical set of
repeated experiments\note{We have tried in this \documentname\ to
  be a bit careful with words.  Following \citet{jaynes}, we have
  tried to use the word ``probability'' for our knowledge or
  uncertainty about a parameter in the problem, and ``frequency'' for
  the rate at which a random variable comes a particular way.  So the
  noise process that sets the uncertainty $\sigma_{yi}$ has a
  frequency distribution, but if we are just talking about our
  \emph{knowledge} about the true value of $y$ for a data point
  measured to have value $y_i$, we might describe that with a
  probability distribution.  In the Bayesian context, then---to get
  extremely technical---the posterior probability distribution
  function is truly a probability distribution, but the likelihood
  factor---really, our generative model---is constructed from
  frequency distributions.}) of getting a value in the infinitesimal
range $[y_i,y_i+\d y]$ per unit $\d y$.

The generative model provides us with a natural, justified, scalar
objective: We seek the line (parameters $m$ and $b$) that maximize the
probability of the observed data given the model or (in standard
parlance) the \emph{likelihood of the parameters}.\note{The likelihood
  $\like$ has been defined in this \documentname\ to be the frequency
  distribution for the observables evaluated at the observed data
  $\ally$.  It is a function of both the data and the model
  parameters.  Although it appears as an explicit function of the
  data, this is called ``the likelihood for the parameters'' and it is
  thought of as a function of the parameters in subsequent inference.
  In some sense, this is because the data are \emph{not} variables but
  rather fixed observed values, and it is only the parameters that are
  permitted to vary.}  In our generative model the data points are
independently drawn (implicitly), so the likelihood $\like$ is the
product of conditional probabilities
\begin{equation}\label{eq:like}
\like = \prod_{i=1}^N \ p(y_i|x_i,\sigma_{yi},m,b) \quad .
\end{equation}
Taking the logarithm,
\begin{eqnarray}\displaystyle
\ln\like
 & = & K - \sum_{i=1}^N \frac{[y_i - m\,x_i - b]^2}{2\,\sigma_{yi}^2} \nonumber\\
 & = & K - \frac{1}{2}\,\chi^2 \quad ,
\end{eqnarray}
where $K$ is some constant.  This shows that likelihood maximization
is identical to $\chi^2$ minimization and we have justified,
scientifically, the procedure of the previous \sectionname.

The Bayesian generalization of this is to say that
\begin{equation}
p(m,b|\ally,I) = \frac{p(\ally|m,b,I)\,p(m,b|I)}{p(\ally|I)} \quad ,
\end{equation}
where $m$ and $b$ are the model parameters, $\ally$ is a short-hand
for all the data $y_i$, $I$ is a short-hand for all the prior
knowledge of the $x_i$ and the $\sigma_{yi}$ and everything else about
the problem\note{One oddity in this \sectionname\ is that when we
  set up the Bayesian problem we put the $\allx$ and $\allsigmay$ into
  the prior information $I$ and \emph{not} into the data.  These were
  not considered data.  Why not?  This was a choice, but we made it to
  emphasize that in the standard approach to fitting, there is a
  \emph{total asymmetry} between the $x_i$ and the $y_i$.  The $x_i$
  are considered part of the \emph{experimental design}; they are the
  input to an experiment that got the $y_i$ as output.  For example,
  the $x_i$ are the times (measured by your perfect and reliable
  clock) at which you chose \apriori\ to look through the telescope,
  and the $y_i$ are the declinations you measured at those times.  As
  soon as there is any sense in which the $x_i$ are themselves
  \emph{also} data, the method---Bayesian or frequentist---of this
  \sectionname\ is invalid.}, $p(m,b|\ally,I)$ is the \emph{posterior}
probability distribution for the parameters given the data and the
prior knowledge, $p(\ally|m,b,I)$ is the likelihood $\like$ just
computed in \equationname~(\ref{eq:like}) (which is a frequency
distribution for the data), $p(m,b|I)$ is the \emph{prior} probability
distribution for the parameters that represents all knowledge
\emph{except} the data $\ally$, and the denominator can---for our
purposes---be thought of as a normalization constant, obtained by
marginalizing the numerator over all parameters. Unless the prior
$p(m,b|I)$ has a lot of variation in the region of interest, the
posterior distribution function $p(m,b|\ally,I)$ here is going to look
very similar to the likelihood function in
\equationname~(\ref{eq:like}) above.  That is, with an uninformative
prior, the straight line that maximizes the likelihood will come
extremely close to maximizing the posterior probability distribution
function; we will return to this later.

We have succeeded in justifying the standard method as optimizing a
justified, scalar objective function; the likelihood of a generative
model for the data.  It is just the great good luck of Gaussian
distributions that the objective can be written as a quadratic
function of the data.  This makes the optimum a linear function of the
data, and therefore trivial to obtain.  This is a miracle.\note{It
  remains a miracle (to us) that the optimization of the $\chi^2$
  objective, which is the only sensible objective under the
  assumptions of the standard problem, has a linear solution.  One can
  attribute this to the magical properties of the Gaussian
  distribution, but the Gaussian distribution is also the
  maximum-entropy distribution (the least informative possible
  distribution) constrained to have zero mean and a known variance; it
  is the limiting distribution of the central limit theorem.  That
  this leads to a convex, linear algebra solution is something for
  which we all ought to give thanks.}

\begin{problem}
Imagine a set of $N$ measurements $t_i$, with uncertainty variances
$\sigma_{ti}^2$, all of the same (unknown) quantity $T$.  Assuming the
generative model that each $t_i$ differs from $T$ by a
Gaussian-distributed offset, taken from a Gaussian with zero mean and
variance $\sigma_{ti}^2$, write down an expression for the log
likelihood $\ln\like$ for the data given the model parameter $T$.
Take a derivative and show that the maximum likelihood value for $T$
is the usual weighted mean.
\end{problem}

\begin{problem}
Take the matrix formulation for $\chi^2$ given in
\equationname~(\ref{eq:chisquared}) and take derivatives to show that
the minimum is at the matrix location given in
\equationname~(\ref{eq:lsf}).
\end{problem}

\section{Pruning outliers}\label{sec:outliers}

The standard linear fitting method is very sensitive to
\emph{outliers}, that is, points that are substantially farther from
the linear relation than expected---or not on the relation at
all---because of unmodeled experimental uncertainty or unmodeled but
rare sources of noise (or because the model has a restricted scope and
doesn't apply to all the data).  There are two general approaches to
mitigating this sensitivity, which are not necessarily different.  The
first is to find ways to objectively remove, reject, or become
insensitive to ``bad'' points; this is the subject of
this \sectionname.  The second is to better model your data-point
uncertainties, permitting larger deviations than the Gaussian noise
estimates; this is the subject of
\sectionname~\ref{sec:non-Gaussian}.  Both of these are strongly
preferable to sorting through the data and rejecting points by hand,
for reasons of subjectivity and irreproducibility that we need not
state.  They are also preferable to the standard (in astrophysics
anyway) procedure known as ``sigma clipping'', which is a procedure
and not the outcome of justifiable modeling.\note{\emph{Sigma
    clipping} is a procedure that involves performing the fit,
  identifying the worst outliers in a $\chi^2$ sense---the points that
  contribute more than some threshold $Q^2$ to the $\chi^2$ sum,
  removing those, fitting again, identifying again, and so on to
  convergence.  This procedure is easy to implement, fast, and
  reliable (provided that the threshold $Q^2$ is set high enough), but
  it has various problems that make it less suitable than methods in
  which the outliers are modeled.  One is that sigma clipping is a
  \emph{procedure} and not the result of optimizing an \emph{objective
    function}.  The procedure does not necessarily optimize any
  objective (let alone any \emph{justifiable} objective).  A second is
  that the procedure gives an answer that depends, in general, on the
  starting point or initialization, and because there is there is no
  way to compare different answers (there is no objective function),
  the investigator can't decide which of two converged answers is
  ``better''.  You might think that the answer with least scatter
  (smallest $\chi^2$ per data point) is best, but that will favor
  solutions that involve rejecting most of the data; you might think
  the answer that uses the most data is best, but that can have a very
  large $\chi^2$.  These issues relate to the fact that the method
  does not explicitly \emph{penalize} the rejection of data; this is
  another bad consequence of not having an explicit objective
  function.  A third problem is that the procedure does not
  necessarily converge to anything non-trivial at all; if the
  threshold $Q^2$ gets very small, there are situations in which all
  but two of the data points can be rejected by the procedure.  All
  that said, with $Q^2\gg 1$ and standard, pretty good data, the
  sigma-clipping procedure is easy to implement and fast; we have
  often used it ourselves.

  Complicating matters is the fact that sigma clipping is often
  employed in the common situation in which you have to perform data
  rejection but you \emph{don't know the magnitudes of your
    uncertainties} $\allsigmay$.  We will say a bit more about this in
  \sectionname~\ref{sec:goodness}, but a better procedure is to
  include the $\allsigmay$ as (possibly restricted) model parameters
  and infer them and marginalize over them \emph{also}.  This is only
  really tractable if you can assume that all the data-point
  uncertainties are identical or you know something else equally
  informative about their \emph{relative} magnitudes.}

If we want to explicitly and objectively reject bad data points, we
must add to the problem a set of $N$ binary integers $q_i$, one per
data point, each of which is unity if the $i$th data point is good,
and zero if the $i$th data point is bad.\note{Outlier modeling of this
  kind was used by \cite{pressH0} to perform a meta-analysis of
  astronomical results, and is discussed at some length in a general
  way by \cite{jaynes}.}  In addition, to construct an objective
function, one needs a parameter $\Pbad$, the \emph{prior} probability
that any individual data point is bad, and parameters $(\Ybad,\Vbad)$,
the mean and variance of the distribution of bad points (in $y$).  We
need these latter parameters because, as we have emphasized above, our
inference comes from evaluating a probabilistic generative model for
the data, and that model must generate the bad points as well as the
good points!

All these $N+3$ extra parameters $(\allq,\Pbad,\Ybad,\Vbad)$ may seem
like crazy baggage, but their values can be \emph{inferred} and
\emph{marginalized out} so in the end, we will be left with an
inference just of the line parameters $(m,b)$.  In general, there is
no reason to be concerned just because you have more parameters than
data.\note{The exponential outlier model has a total of $N+5$
  parameters, with $N$ data points: It has more parameters than data!
  Indeed, some would consider optimization or inference in this
  situation to be impossible.  Of course it is not impossible, and for
  reasons that are instructive, though any detailed discussion is
  beyond the scope of this \documentname.  Fundamentally, optimization
  of this model is possible only when there are \emph{informative
    priors}; in this case the informative prior is that all of the
  badness bits $\allq$ are integers, and every one of them can take on
  only the value zero or unity.  This is a very strong prior, and
  limits the amount of information that can flow from the parameters
  of interest ($m$ and $b$) into the uninteresting parameters $\allq$.
  More generally, there is no limit on the \emph{number} of parameters
  that can be constrained with a given data set; there is only a limit
  on the amount of \emph{information} that those parameters can carry
  or obtain from the data.  There is an additional subtlety, which
  goes way beyond the scope of this document, which is that the
  marginalization over the uninteresting parameters integrates out and
  thereby reduces the degeneracies.}  However, the marginalization
will require that we have a \emph{measure} on our parameters
(integrals require measures) and that measure is provided by a prior.
That is, the marginalization will require, technically, that we become
Bayesian; more on this below.

In this case, the likelihood is
\begin{eqnarray}\displaystyle
\like &\equiv& p(\ally|m,b,\allq,\Ybad,\Vbad,I)
 \nonumber\\
\like &=& \prod_{i=1}^N
 \left[\pfg(\ally|m,b,I))\right]^{q_i}\,
 \left[\pbg(\ally|\Ybad,\Vbad,I)\right]^{[1-q_i]}
 \nonumber\\
\like &=& \prod_{i=1}^N \left[\frac{1}{\sqrt{2\,\pi\,\sigma_{yi}^2}}
 \,\exp\left(-\frac{[y_i-m\,x_i-b]^2}{2\,\sigma_{yi}^2}\right)\right]^{q_i}
 \nonumber \\ & & \quad\times
 \left[\frac{1}{\sqrt{2\,\pi\,[\Vbad+\sigma_{yi}^2]}}
 \,\exp\left(-\frac{[y_i-\Ybad]^2}{2\,[\Vbad+\sigma_{yi}^2]}\right)\right]^{[1-q_i]}
 \quad ,
\end{eqnarray}
where $\pfg(\cdot)$ and $\pbg(\cdot)$ are the generative models for
the foreground (good, or straight-line) and background (bad, or
outlier) points.  Because we are permitting data rejection, there is
an important prior probability on the $\allq$ that penalizes each
rejection:
\begin{eqnarray}\displaystyle
p(m,b,\allq,\Pbad,\Ybad,\Vbad|I)
 &=& p(\allq|\Pbad,I)\,p(m,b,\Pbad,\Ybad,\Vbad|I)
 \nonumber\\
p(\allq|\Pbad,I)
 &=& \prod_{i=1}^N[1-\Pbad]^{q_i}\,\Pbad^{[1-q_i]}
 \quad ,
\end{eqnarray}
that is, the binomial probability of the particular sequence $\allq$.
The priors on the other parameters $(\Pbad,\Ybad,\Vbad)$ should either
be set according to an analysis of your prior knowledge or else be
uninformative (as in flat in $\Pbad$ in the range $[0,1]$, flat in
$\Ybad$ in some reasonable range, and flat in the logarithm of $\Vbad$
in some reasonable range).  If your data are good, it won't matter
whether you choose uninformative priors or priors that really
represent your prior knowledge; in the end good data dominate either.

We have made a somewhat restrictive assumption that all data points
are equally likely \apriori\ to be bad, but you rarely know enough
about the badness of your data to assume anything better.\note{When
  flagging bad data, you might not want to give all points the same
  prior probability distribution function over $\Pbad$, for example,
  when you combine data sets from different sources.  Of course it is
  rare that one has \emph{reliable} information about the
  \emph{unreliability} of one's data.}  The fact that the prior
probabilities $\Pbad$ are all set equal, however, does not mean that
the \emph{posterior} probabilities that individual points are bad will
be at all similar.  Indeed, this model permits an objective ranking
(and hence public embarassment) of different contributions to any data
set.\note{One of the amusing things about the Bayesian posterior
  for the data rejections $\allq$ is that you can pick a particular
  data point $J$ and marginalize over $(m,b,\Pbad,\Ybad,\Vbad)$ and
  all the $q_i$ \emph{except} $q_J$.  This will return the
  marginalized posterior probability that point $J$ is good.  This is
  valuable for embarassing colleagues in meta-analyses
  \citep{pressH0}.  Indeed, if embarassing colleagues is your goal,
  then Bayesian methods for data rejection are useful: They permit
  parameter-independent (that is, marginalized) statements about the
  probabilities that individual data points are bad.  This is possible
  even in the marginalized mixture model: For any data point, at any
  setting of the parameters $(m,b,\Pbad,\Ybad,\Vbad)$, the relative
  amplitude of the good and bad parts of the generative frequency
  distributions in equation~(\ref{eq:mixture}) is the odds that the
  data point is bad (or good, depending on the direction of the ratio
  and one's definition of the word ``odds'').  It is possible
  therefore to marginalize over \emph{all} the parameters, obtaining
  the mean, for each data point $i$, of the foreground frequency
  $\left<[1-\Pbad]\,\pfg(\cdot)\right>$ (mean taken over all samples
  from the posterior) evaluated at the data point and of the
  background frequency $\left<\Pbad\,\pbg(\cdot)\right>$.  The ratio
  of these is the marginalized odds that data point $i$ is bad.}

We have used in this likelihood formulation a Gaussian model for the
bad points; is that permitted?  Well, we don't know much about the bad
points (that is one of the things that makes them bad), so this
Gaussian model must be wrong in detail.  But the power of this and the
methods to follow comes not from making an \emph{accurate} model of
the outliers, it comes simply from \emph{modeling} them.  If you have
an accurate or substantiated model for your bad data, use it by all
means.  If you don't, because the Gaussian is the maximum-entropy
distribution described by a mean and a variance, it is---in
some sense---the \emph{least} restrictive assumption, given that we
have allowed that mean and variance to vary in the fitting (and,
indeed, we will marginalize over both in the end).

The posterior probability distribution function, in general, is the
likelihood times the prior, properly normalized.  In this case, the
posterior we (tend to) care about is that for the line parameters
$(m,b)$.  This is the full posterior probability distribution
marginalized over the bad-data parameters $(\allq,\Pbad,\Ybad,\Vbad)$.
Defining $\pvtheta\equiv(m,b,\allq,\Pbad,\Ybad,\Vbad)$ for brevity, the
full posterior is
\begin{equation}
p(\pvtheta,I) =
 \frac{p(\ally|\pvtheta,I)}{p(\ally|I)}
 \,p(\pvtheta|I)
 \quad,
\end{equation}
where the denominator can---for our purposes here---be thought of as a
normalization integral that makes the posterior distribution function
integrate to unity.  The marginalization looks like
\begin{equation}
p(m,b|\ally,I)=\int \d\allq\,\d\Pbad\,\d\Ybad\,\d\Vbad
 \,p(\pvtheta,I) \quad,
\end{equation}
where by the integral over $\d\allq$ we really mean a sum over all of
the $2^N$ possible settings of the $q_i$, and the integrals over the
other parameters are over their entire prior support.  Effectively,
then, this marginalization involves evaluating $2^N$ different
likelihoods and marginalizing each one of them over
$(\Pbad,\Ybad,\Vbad)$!  Of course, because very few points are true
candidates for rejection, there are many ways to ``trim the tree'' and
do this more quickly, but thought is not necessary: Marginalization is
in fact analytic.

Marginalization of this---what we will call ``exponential''---model
leaves a much more tractable---what we will call ``mixture''---model.
Imagine marginalizing over an individual $q_i$.  After this
marginalization, the $i$th point can be thought of as being drawn from
a \emph{mixture} (sum with amplitudes that sum to unity) of the
straight-line and outlier distributions.  That is, the generative
model for the data $\ally$ integrates (or sums, really) to
\begin{eqnarray}\label{eq:mixture}\displaystyle
\like &\equiv& p(\ally|m,b,\Pbad,\Ybad,\Vbad,I)
 \nonumber\\
\like &\equiv& \prod_{i=1}^N
 \left[ (1-\Pbad)\,\pfg(\ally|m,b,I))
 + \Pbad\,\pbg(\ally|\Ybad,\Vbad,I) \right]
 \nonumber\\
\like &\propto&
 \prod_{i=1}^N \left[\frac{1-\Pbad}{\sqrt{2\,\pi\,\sigma_{yi}^2}}
 \,\exp\left(-\frac{[y_i-m\,x_i-b]^2}{2\,\sigma_{yi}^2}\right)\right.
 \nonumber \\ & & \quad
 \left.+ \frac{\Pbad}{\sqrt{2\,\pi\,[\Vbad+\sigma_{yi}^2]}}
 \,\exp\left(-\frac{[y_i-\Ybad]^2}{2\,[\Vbad+\sigma_{yi}^2]}\right)\right]
 \quad ,
\end{eqnarray}
where $\pfg(\cdot)$ and $\pbg(\cdot)$ are the generative models for
the foreground and background points, $\Pbad$ is the probability that
a data point is bad (or, more properly, the amplitude of the bad-data
distribution function in the mixture), and $(\Ybad,\Vbad)$ are
parameters of the bad-data distribution.

Marginalization of the posterior probability distribution function
generated by the product of the mixture likelihood in
\equationname~(\ref{eq:mixture}) times a prior on
$(m,b,\Pbad,\Ybad,\Vbad)$ does \emph{not} involve exploring an
exponential number of alternatives.  An integral over only three
parameters produces the marginalized posterior probability
distribution function for the straight-line parameters $(m,b)$.  This
scale of problem is very well-suited to simple multi-dimensional
integrators, in particular it works very well with a simple
Markov-Chain Monte Carlo, which we advocate below.

We have lots to discuss.  We have gone from a procedure
(\sectionname~\ref{sec:standard}) to an objective function
(\sectionname~\ref{sec:objective}) to a general expression for a
non-trivial posterior probability distribution function involving
priors and multi-dimensional integration.  We will treat each of
these issues in turn: We must set the prior, we must find a way to
integrate to marginalize the distribution function, and then we have
to decide what to do with the marginalized posterior: Optimize it or
sample from it?

It is unfortunate that prior probabilities must be specified.
Conventional frequentists take this problem to be so severe that for
some it invalidates Bayesian approaches entirely!  But a prior is a
\emph{necessary condition} for marginalization, and marginalization is
necessary whenever extra parameters are introduced beyond the two
$(m,b)$ that we care about for our straight-line fit.\note{The
  likelihood---for example, that in equation~(\ref{eq:mixture})---is
  the likelihood ``for the parameters'' but it has units (or
  dimensions), in some sense, of inverse data, because it is computed
  as the probability distribution function for observations, evaluated
  at the observed data.  The likelihood, therefore, is not something
  you can marginalize---the integral of the likelihood over a
  \emph{parameter} has incomprehensible units.  The posterior
  probability distribution function, on the other hand, has units (or
  dimensions) of inverse parameters, so it can be integrated over the
  parameter space.  Marginalization, therefore, is only possible after
  construction of a posterior probability distribution function;
  marginalization is only possible in Bayesian analyses.  This is
  related to the fact that the prior probability distribution function
  serves (in part) to define a measure in the parameter space.  You
  can't integrate without a measure.

  If all you care about is parameters $(m,b)$, and, furthermore, all
  you care about is the MAP answer, you must chose the MAP values from
  the \emph{marginalized} posterior.  This is because the MAP value of
  $(m,b,\Pbad,\Ybad,\Vbad)$ in the unmarginalized posterior will not,
  in general, have the same value of $(m,b)$ as the MAP value in the
  marginalized posterior.  This point is important in many real cases
  of inference: Even small departures from Gaussianity in the
  posterior distribution function will lead to substantial shifts of
  the mode under marginalization.  Of course, if what you are carrying
  forward is not the MAP result but a \emph{sampling} from the
  posterior, a sampling from the unmarginalized posterior will contain
  the same distribution of $(m,b)$ values as a sampling from the
  marginalized one, by definition.

  This \documentname\ is not an argument in favor of Bayesian
  approaches; sometimes non-Bayesian approaches are much more
  expedient.  What this \documentname\ argues for is \emph{generative
    modeling}, which \emph{is} demanding.  The standard objection to
  Bayesian methods is that they require priors, and those priors are
  hard (some say impossible) to set objectively.  But this
  difficulty---in most problems of interest---pales in comparison to
  the difficulty of writing down a justifiable or even approximate
  generative model for the data.

  While we claim to be neutral on Bayesianism relative to frequentism,
  there is one very important respect in which frequentists are at a
  disadvantage when there are nuisance parameters.  If the nuisance
  parameters are powerful, and there are acceptable fits to the data
  that show a large range in the parameters of interest---as there
  will be in these mixture models when $\Pbad$ is made large---then a
  frequentist, who is not permitted to marginalize, will be able to
  conclude nothing interesting about the parameters of interest
  anywhere other than at the maximum-likelihood point.  That is, the
  maximum-likelihood point might have a small $\Pbad$ and very good
  results for line parameters $(m,b)$, but if the frequentist wants to
  be responsible and report the full range of models that is
  \emph{acceptable} given the data, the frequentist must report that
  the data are also consistent with almost all the data being bad
  ($\Pbad\approx 1$) and just about any line parameters you like.
  Marginalization stops this, because the specificity of the data
  explanation under small $\Pbad$ permits the small-$\Pbad$ regions of
  parameter space to dominate the marginalization integral.  But if
  you have no measure (no prior), you can perform no marginalization,
  and the responsible party may be forced to report that no useful
  confidence interval on the line parameters are possible in this
  mixture model.  That is a perfectly reasonable position, but
  demonstrates the cost of strict adherence to frequentism (which, for
  our purposes, is the position that there is no believable or natural
  measure or prior on the parameters).}  Furthermore, in any situation
of straight-line fitting where the data are numerous and generally
good, these priors do not matter very much to the final answer.  The
way we have written our likelihoods, the peak in the posterior
distribution function becomes very directly analogous to what you
might call the maximum-likelihood answer when the (improper,
non-normalizable) prior ``flat in $m$, flat in $b$'' is adopted, and
something sensible has been chosen for $(\Pbad,\Ybad,\Vbad)$.  Usually
the investigator \emph{does} have some basis for setting some more
informative priors, but a deep and sensible discussion of priors is
beyond the scope of this \documentname.  Suffice it to say, in this
situation---pruning bad data---the setting of priors is the
\emph{least} of one's problems.

The second unfortunate thing about our problem is that we must
marginalize.  It is also a \emph{fortunate} thing that we \emph{can}
marginalize.  As we have noted, marginalization is necessary if you
want to get estimates with uncertainties or confidence intervals in
the straight-line parameters $(m,b)$ that properly account for their
covariances with the nuisance parameters $(\Pbad,\Ybad,\Vbad)$.  This
marginalization can be performed by direct numerical integration
(think gridding up the nuisance parameters, evaluating the posterior
probability at every grid point, and summing), or it can be performed
with some kind of sampler, such as a Monte-Carlo Markov Chain.  We
strongly recommend the latter because in situations where the integral
is of a \emph{probability distribution}, not only does the MCMC scheme
integrate it, it also produces a \emph{sampling} from the distribution
as a side-product for free.

The standard Metropolis--Hastings MCMC procedure works extremely well
in this problem for both the optimization and the integration and
sampling.  A discussion of this algorithm is beyond the scope of this
document, but briefly the algorithm is this simple: Starting at some
position in parameter space, where you know the posterior probability,
iterate the following: \textsl{(a)}~Step randomly in parameter
space. \textsl{(b)}~Compute the new posterior probability at the new
position in parameter space. \textsl{(c)}~Draw a random number $R$ in
the range $0<R<1$.  \textsl{(d)}~If $R$ is less than the ratio of the
new posterior probability to the old (pre-step) probability, accept
the step, add the new parameters to the chain; if it is greater,
reject the step, re-add the old parameters to the chain.  The devil is
in the details, but good descriptions abound\note{There is a good
  introduction to Metropolis--Hastings MCMC in \citealt{press}, among
  other places.} and that with a small amount of experimentation (try
Gaussians for the proposal distribution and scale them until the
acceptance ratio is about half) rapid convergence and sampling is
possible in this problem. A large class of MCMC algorithms exists,
many of which outperform the simple Metropolis algorithm with little
extra effort.\note{Variations of MCMC that go beyond
  Metropolis--Hastings---and the conditions under such variations make
  sense---are described in, for example, \cite{gilksmcmc},
  \cite{neal2003a}, and \cite{mackay}.}

Once you have run an MCMC or equivalent and obtained a chain of
samples from the posterior probability distribution for the full
parameter set $(m,b,\Pbad,\Ybad,\Vbad)$, you still must decide what to
report about this sampling.  A histogram of the samples in any one
dimension is an approximation to the fully marginalized posterior
probability distribution for that parameter, and in any two is that
for the pair.  You can find the ``maximum \aposteriori'' (or MAP)
value, by taking the peak of the marginalized posterior probability
distribution for the line parameters $(m,b)$.  This, in general, will
be different from the $(m,b)$ value at the peak of the unmarginalized
probability distribution for \emph{all} the parameters
$(m,b,\Pbad,\Ybad,\Vbad)$, because the posterior distribution can be
very complex in the five-dimensional space.  Because, for our
purposes, the three parameters $(\Pbad,\Ybad,\Vbad)$ are nuisance
parameters, it is more correct to take the ``best-fit'' $(m,b)$ from
the marginalized distribution.

Of course finding the MAP value from a sampling is not trivial.  At
best, you can take the highest-probability sample.  However, this
works only for the unmarginalized distribution.  The MAP value from
the marginalized posterior probability distribution function for the
two line parameters $(m,b)$ will not be the value of $(m,b)$ for the
highest-probability sample.  To find the MAP value in the marginalized
distribution, it will be necessary to make a histogram or perform
density estimation in the sample; this brings up an enormous host of
issues way beyond the scope of this \documentname.  For this reason,
we usually advise just taking the mean or median or some other simple
statistic on the sampling.  It is also possible to establish
confidence intervals (or credible intervals) using the distribution of
the samples.\note{Given a sampling of the posterior probability
  distribution function or the likelihood function of the parameters,
  it is most responsible to report---if not that entire
  sampling---some confidence intervals or credible intervals.  There
  is a terminology issue in which the term ``confidence interval'' has
  become associated with frequentist statistics, which is unfortunate
  because Bayesian statistics is the calculus of plausibility or
  confidence.  But whether you call them confidence intervals or
  credible intervals, the standard way to produce the confidence
  interval of fractional size $f$ ($f=0.95$ for the 95~percent
  interval) on, say, your parameter $m$, is as follows: Rank the
  samples in order of increasing $m$ and take either the smallest
  interval that contains a fraction $f$ of them, or else the interval
  which excludes the first and last $[1-f]/2$ of them.  We usually do
  the latter, since it is fast and easy to explain.  Then a good value
  to report is the median of the sampling, and quantiles around that
  median.  But it does not matter much so long as the report is clear
  and sensible.}

While the MAP answer---or any other simple ``best answer''---is
interesting, it is worth keeping in mind the third unfortunate thing
about any Bayesian method: It does not return an ``answer'' but rather
it returns a \emph{posterior probability distribution}.  Strictly,
this posterior distribution function \emph{is} your answer.  However,
what scientists are usually doing is not inference but rather
\emph{decision-making}.  That is, the investigator wants a specific
answer, not a distribution.  There are (at least) two reactions to
this.  One is to ignore the fundamental Bayesianism at the end, and
choose simply the MAP answer.  This is the Bayesian's analog of
maximum likelihood; the standard uncertainty on the MAP answer would
be based on the peak curvature or variance of the marginalized
posterior distribution.  The other reaction is to suck it up and
sample the posterior probability distribution and carry forward not
one answer to the problem but $M$ answers, each of which is drawn
fairly and independently from the posterior distribution function.
The latter is to be preferred because \textsl{(a)}~it shows the
uncertainties very clearly, and \textsl{(b)}~the sampling can be
carried forward to future inferences as an approximation to the
posterior distribution function, useful for propagating uncertainty,
or standing in as a \emph{prior} for some subsequent
inference.\note{Posterior probability distribution functions and
samplings thereof are useful for propagating probabilistic information
about a scientific result.  However, they must be used with care: If
the subsequent data analyzer has a prior that overlaps or conflicts
with the prior used in making the posterior sampling, he or she will
need access not just to the posterior probability distribution
function but to the likelihood function.  It is the likelihood
function that contains all the information about the data---all the
new information generated by the experiment---and therefore it is the
likelihood that is most useful to subsequent users.  Even a posterior
sampling should be passed forward with likelihood or prior tags so
that the prior can be divided out or replaced.  Otherwise subsequent
users live in danger of squaring (or worse) the prior.} It is also
returned, trivially (as we noted above) by the MCMC integrator we
advised for performing the marginalization.

\begin{figure}[htbp]
\exampleplottwo{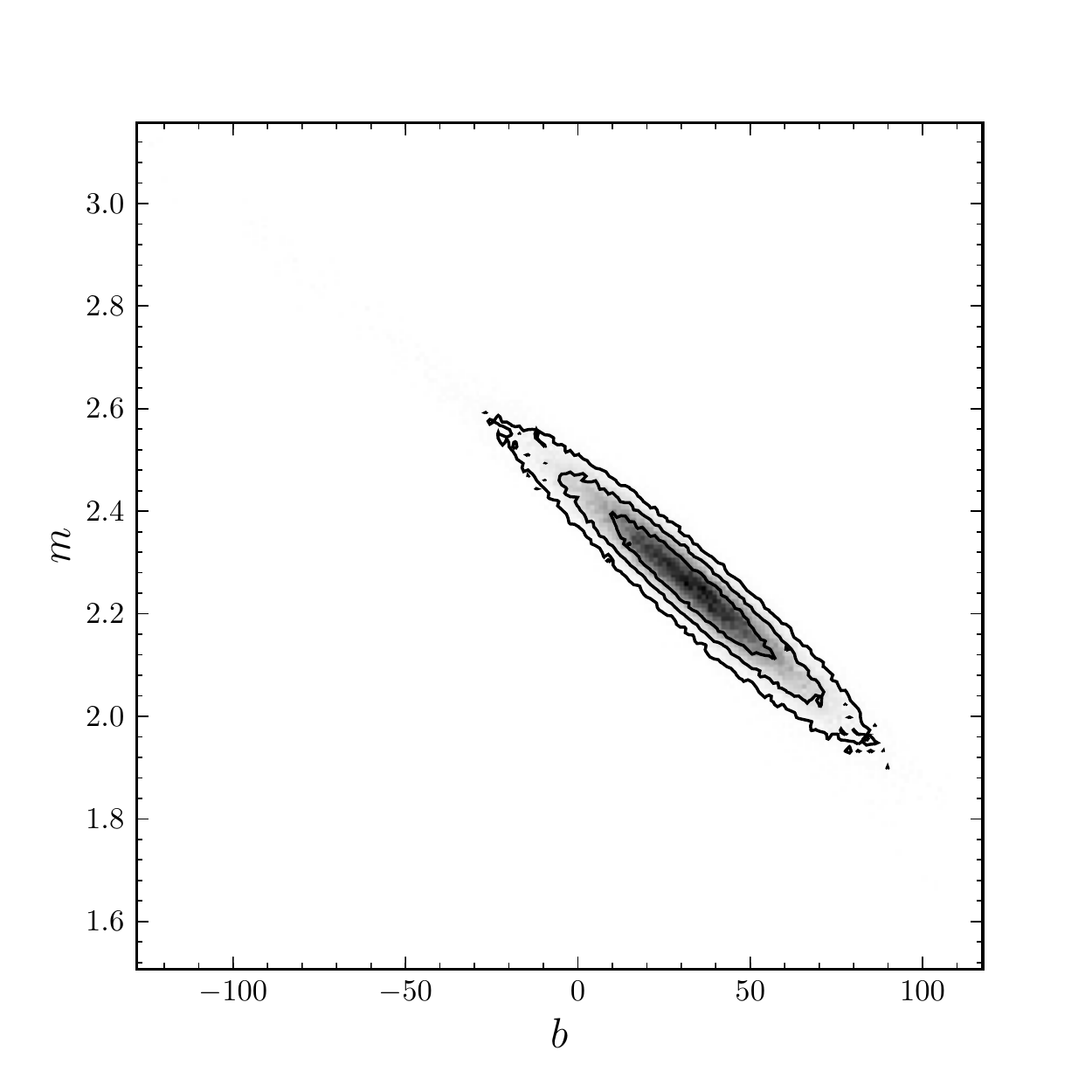}{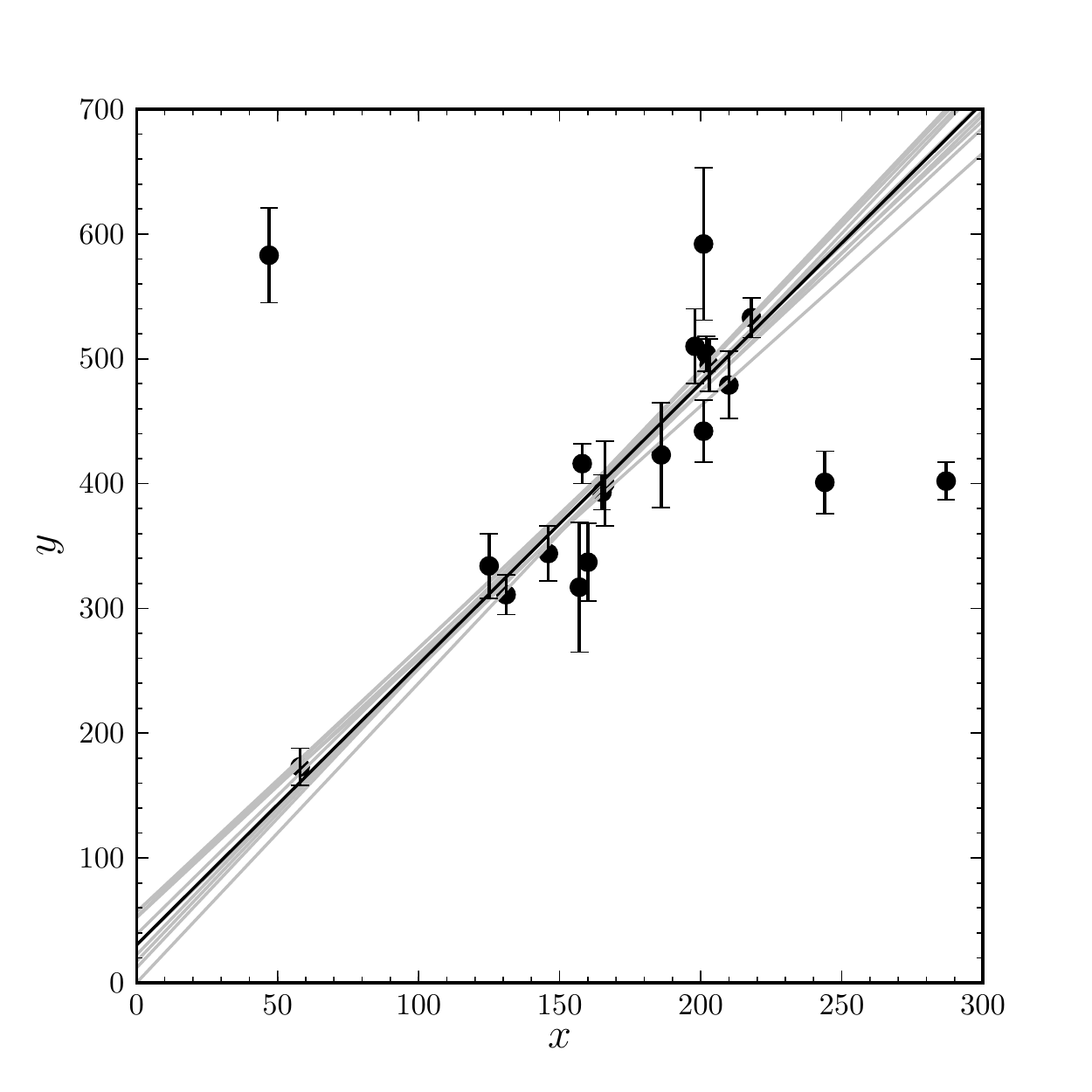}
\caption{Partial solution to \problemname~\ref{prob:mixture}: On the
left, a sampling approximation to the marginalized posterior
probability distribution (left) for the outlier (mixture) model.  On
the right, the marginalized MAP line (dark grey line) and a draw from
the sampling (light grey lines).}\label{fig:mixture}
\end{figure}

\begin{problem}\label{prob:mixture}
Using the mixture model proposed above---that treats the distribution
as a mixture of a thin line containing a fraction $[1-\Pbad]$ of the
points and a broader Gaussian containing a fraction $\Pbad$ of the
points---find the best-fit (the maximum \aposteriori) straight line
$y=m\,x+b$ for the $x$, $y$, and $\sigma_y$ for the data in
\tablename~\ref{table:data_allerr} on
page~\pageref{table:data_allerr}.  Before choosing the MAP line,
marginalize over parameters $(\Pbad,\Ybad,\Vbad)$.  That is, if you
take a sampling approach, this means sampling the full
five-dimensional parameter space but then choosing the peak value in
the histogram of samples in the two-dimensional parameter space
$(m,b)$.  Make one plot showing this two-dimensional histogram, and
another showing the points, their uncertainties, and the MAP line.
How does this compare to the standard result you obtained in
\problemname~\ref{prob:standard}?  Do you like the MAP line better or
worse?  For extra credit, plot a sampling of 10 lines drawn from the
marginalized posterior distribution for $(m,b)$ (marginalized over
$\Pbad,\Ybad,\Vbad$) and plot the samples as a set of light grey or
transparent lines.  Your plot should look like
\figurename~\ref{fig:mixture}.
\end{problem}

\begin{figure}[htbp]
\exampleplottwo{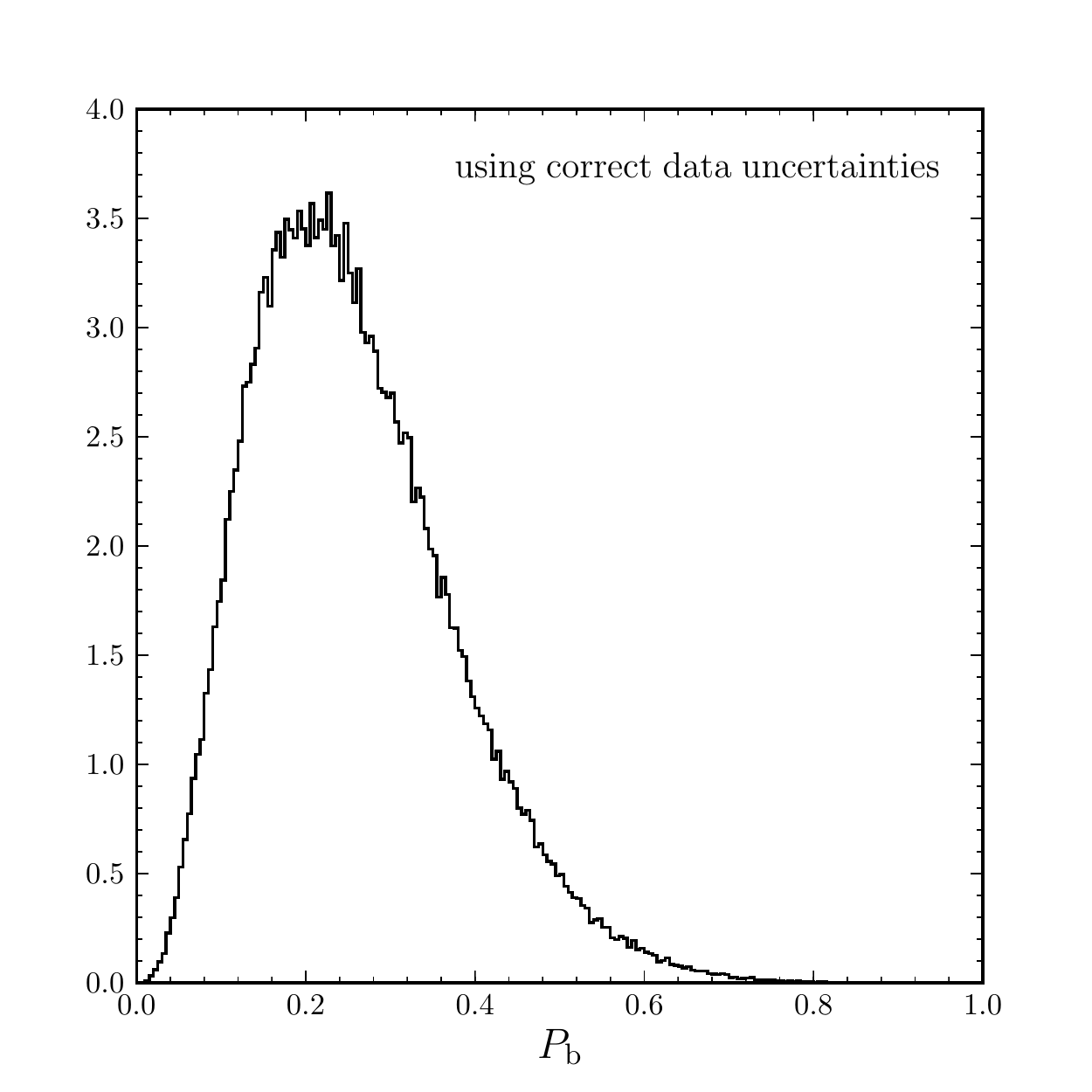}{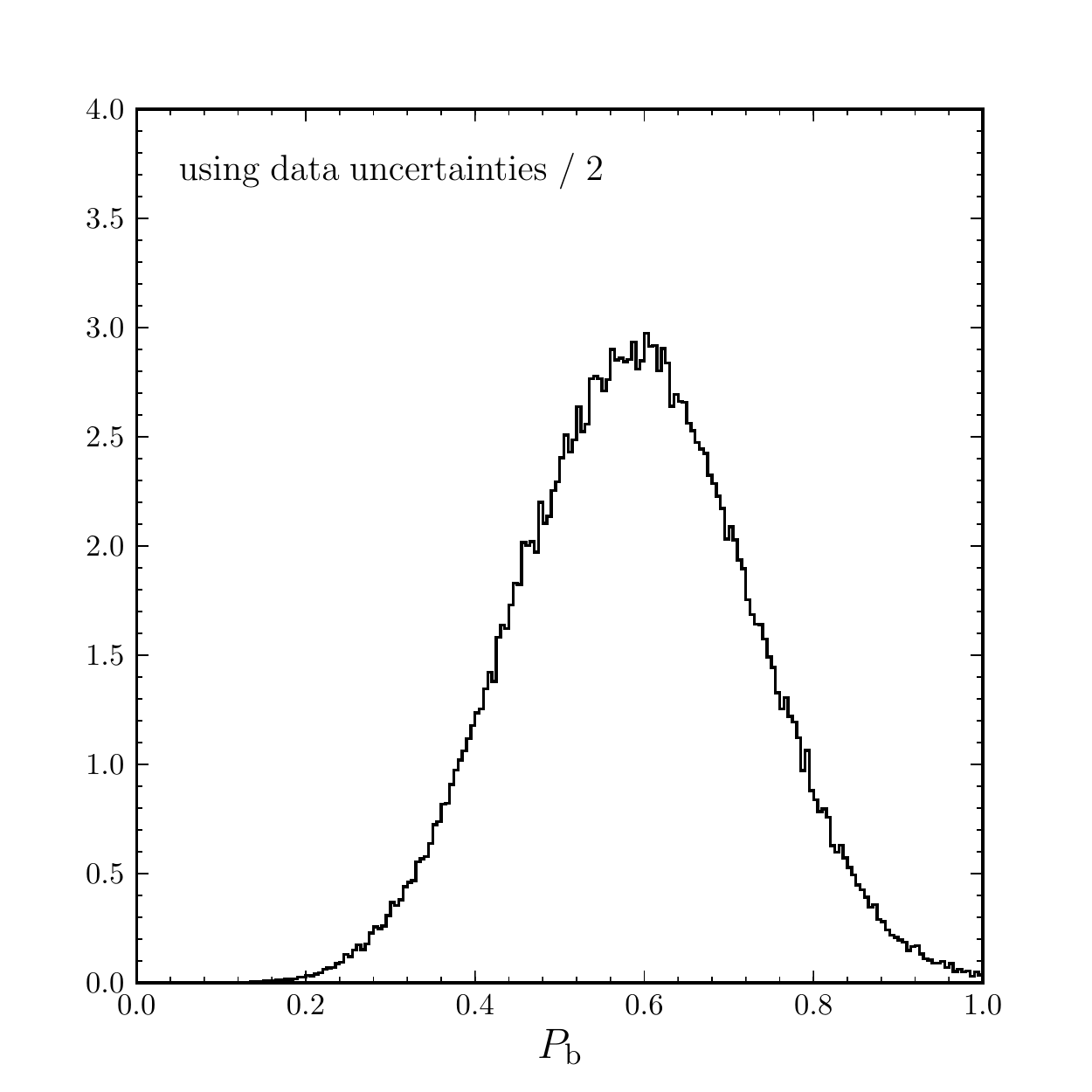}
\caption{Partial solution to \problemname~\ref{prob:badfraction}: The
marginalized posterior probability distribution function for the
$\Pbad$ parameter (prior probability that a point is an outlier); this
distribution gets much worse when the data uncertainties are
underestimated.}\label{fig:badfraction}
\end{figure}

\begin{problem}\label{prob:badfraction}
Solve \problemname~\ref{prob:mixture} but now plot the fully
marginalized (over $m,b,\Ybad,\Vbad$) posterior distribution function
for parameter $\Pbad$.  Is this distribution peaked about where you
would expect, given the data?  Now repeat the problem, but dividing
all the data uncertainty variances $\sigma_{yi}^2$ by 4 (or dividing
the uncertainties $\sigma_{yi}$ by 2).  Again plot the fully
marginalized posterior distribution function for parameter $\Pbad$.
Your plots should look something like those in
\figurename~\ref{fig:badfraction}.  Discuss.
\end{problem}

\section{Uncertainties in the best-fit parameters}\label{sec:uncertainty}

In the standard linear-algebra method of $\chi^2$ minimization given
in \sectionname~\ref{sec:standard}, the uncertainties in the best-fit
parameters $(m,b)$ are given by the two-dimensional output covariance
matrix
\begin{equation}
\left[\begin{array}{cc}
\sigma_{b}^2 & \sigma_{mb} \\
\sigma_{mb} & \sigma_{m}^2
\end{array}\right] = \inverse{\left[\mAT\,\mCinv\,\mA\right]} \quad ,
\end{equation}
where the ordering is defined by the ordering in matrix $\mA$.  These
uncertainties for the model parameters \emph{only} strictly hold under
three extremely strict conditions, none of which is met in most real
situations: \textsl{(a)}~The uncertainties in the data points must
have variances correctly described by the $\sigma_{yi}^2$;
\textsl{(b)}~there must be no rejection of any data or any departure
from the exact, standard definition of $\chi^2$ given in
\equationname~(\ref{eq:chisquared}); and \textsl{(c)}~the generative
model of the data implied by the method---that is, that the data are
truly drawn from a negligible-scatter linear relationship and
subsequently had noise added, where the noise offsets were generated
by a Gaussian process---must be an accurate description of the data.

These conditions are rarely met in practice.  Often the noise
estimates are rough (or missing entirely!), the investigator has
applied data rejection or equivalent conditioning, and the
relationship has intrinsic scatter and curvature.\note{If you are
  sigma clipping or doing anything that looks like standard
  least-square fitting but with small modifications, it might be
  tempting to use the standard uncertainty estimate
  $\inverse{\left[\mAT\,\mCinv\,\mA\right]}$.  But that is definitely
  wrong, because whatever encouraged you to do the sigma clipping or
  to make other modifications is probably sufficient reason to
  disbelieve the standard uncertainty estimate.  In these cases you
  want either to be measuring the width of your posterior probability
  distribution function (if you are a Bayesian) or else using an
  empirical estimate of the uncertainty such as jackknife or
  bootstrap.

  Bayesians like to be smug here---and have some
  justification---because the posterior distribution function gives
  the best fit parameters and the full distribution around that
  best-fit point in parameter space.  However, the smugness must be
  tempered by the fact that \emph{within} the context of a single
  model (such as the generative model described in
  \sectionname~\ref{sec:objective}), the Bayesian analysis does not
  return any useful goodness-of-fit information; a bad model can be
  constrained well by the data but still be dead wrong.}  For these
generic reasons, we much prefer empirical estimates of the uncertainty
in the best-fit parameters.

In the Bayesian outlier-modeling schemes of
\sectionname~\ref{sec:outliers}, the output is a
posterior distribution for the parameters $(m,b)$.  This distribution
function is closer than the standard estimate
$\inverse{\left[\mAT\,\mCinv\,\mA\right]}$ to being an
empirical measurement of the uncertainties of the parameters.  The
uncertainty variances $(\sigma_m^2,\sigma_b^2)$ and the covariance
$\sigma_{mb}$ can be computed as second moments of this posterior
distribution function.  Computing the variances this way does involve
assumptions, but it is not extremely sensitive to the assumption that
the model is a good fit to the data; that is, as the model becomes a
bad fit to the data (for example when the data points are not
consistent with being drawn from a narrow, straight line), these
uncertainties change in accordance.  That is in strong contrast to the
elements of the matrix $\inverse{\left[\mAT\,\mCinv\,\mA\right]}$,
which don't depend in any way on the quality of fit.

\begin{figure}[htbp]
\exampleplottwo{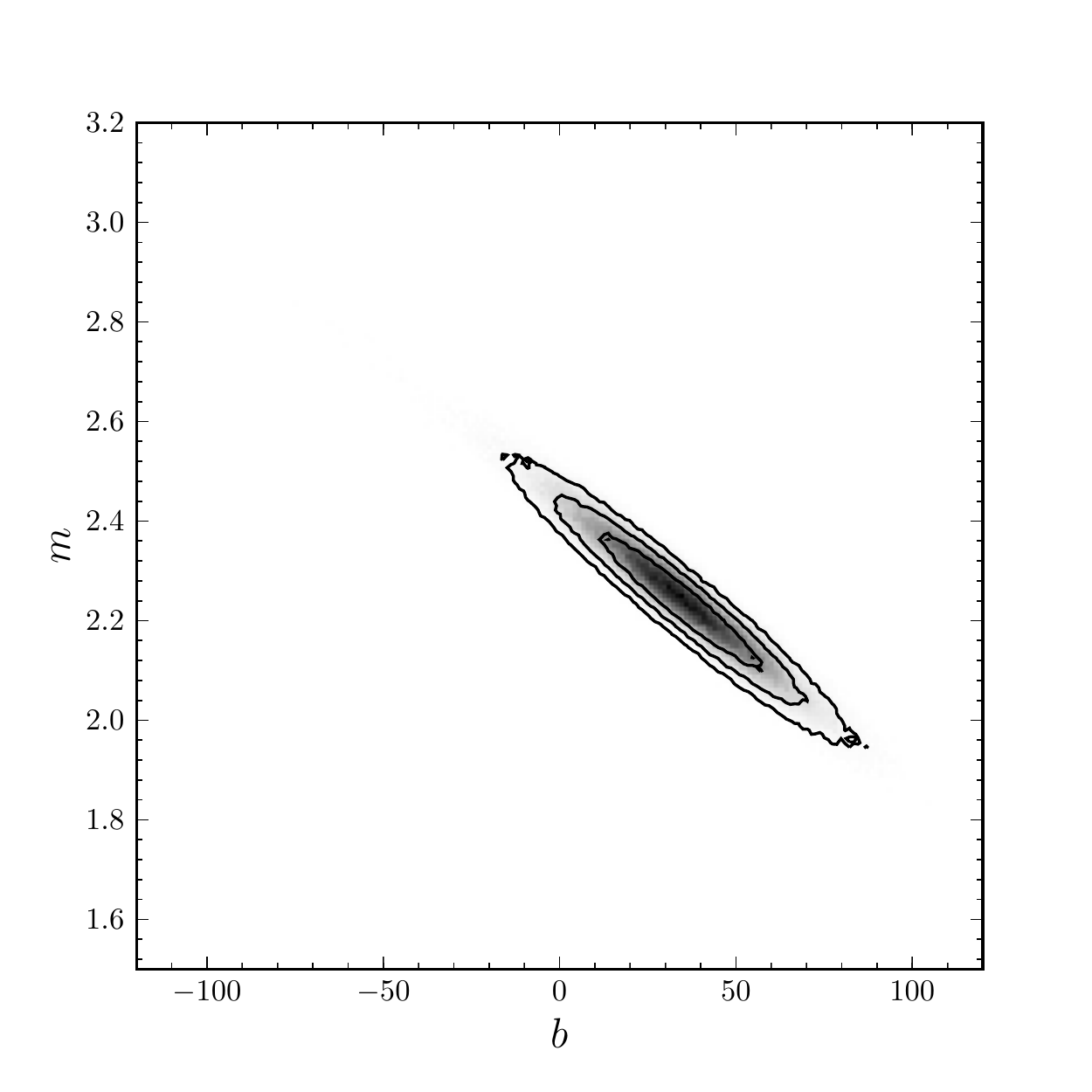}{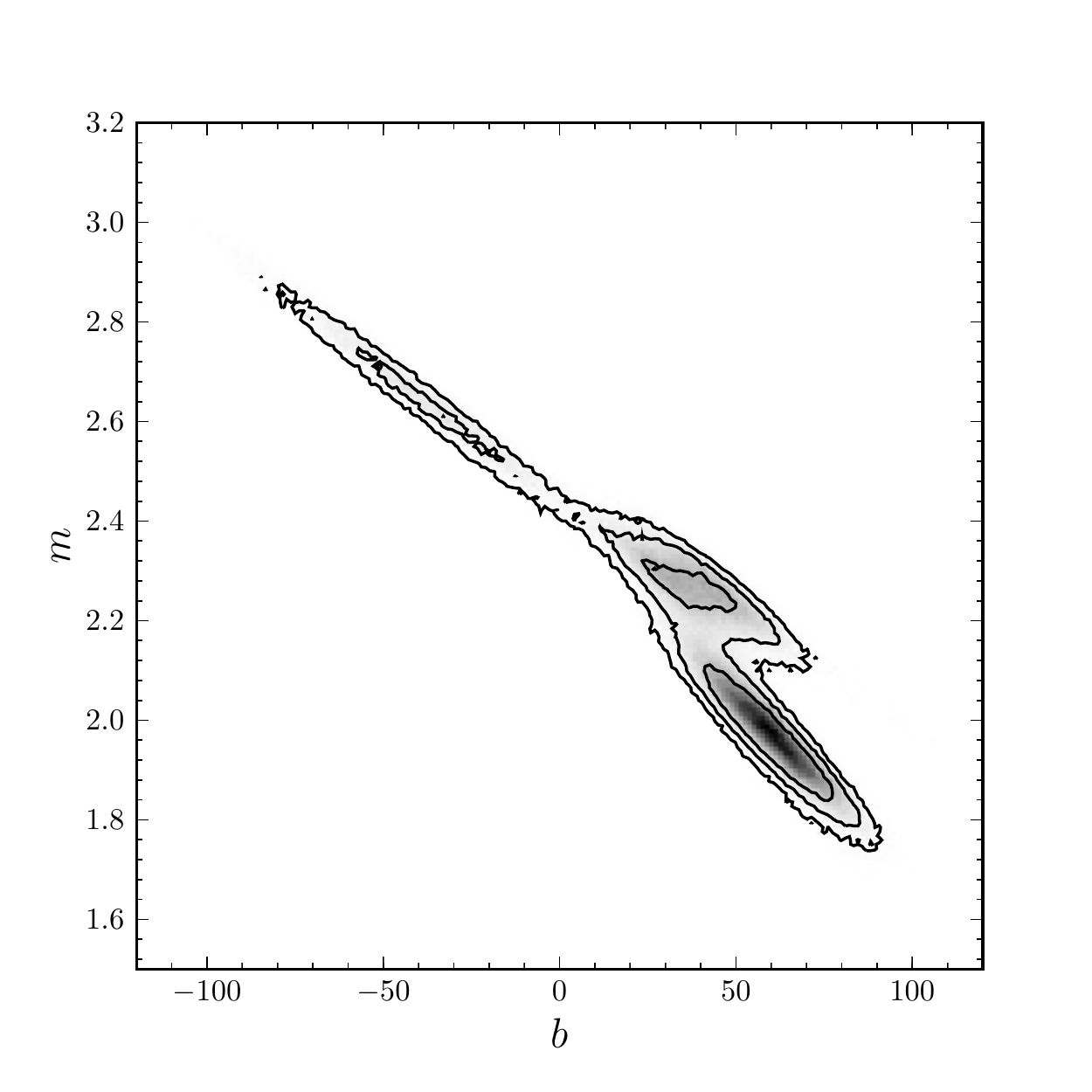}
\caption{Partial solution to \problemname~\ref{prob:reduceerror}: The
posterior probability distribution can become multi-modal (right
panel) when you underestimate your observational
uncertainties.}\label{fig:reduceerror}
\end{figure}

Either way, unless the output of the fit is being used \emph{only} to
estimate the slope, and nothing else, it is a mistake to ignore the
off-diagonal terms of the uncertainty variance tensor.  That is, you
generally know some linear combinations of $m$ and $b$ much, much
better than you know either individually (see, for example,
\figurenames~\ref{fig:mixture} and \ref{fig:reduceerror}).  When the
data are good, this covariance is related to the location of the data
relative to the origin, but any non-trivial propagation of the
best-fit results requires use of either the full $2\times 2$
uncertainty tensor or else a two-dimensional sampling of the
two-dimensional (marginalized, perhaps) posterior distribution.

In any non-Bayesian scheme, or when the full posterior has been
discarded in favor of only the MAP value, or when data-point
uncertainties are not trusted, there are still empirical methods for
determining the uncertainties in the best-fit parameters.  The two
most common are \emph{bootstrap} and \emph{jackknife}.  The first
attempts to empirically create new data sets that are similar to your
actual data set.  The second measures your differential sensitivity to
each individual data point.

In bootstrap, you do the unlikely-sounding thing of drawing $N$ data
points randomly from the $N$ data points you have \emph{with
  replacement}.  That is, some data points get dropped, and some get
doubled (or even tripled), but the important thing is that you select
each of the $N$ that you are going to use in each trial independently
from the whole set of $N$ you have.  You do this selection of $N$
points once for each of $M$ bootstrap trials $j$.  For each of the $M$
trials $j$, you get an estimate---by whatever method you are using
(linear fitting, fitting with rejection, optimizing some custom
objective function)---of the parameters $(m_j,b_j)$, where $j$ goes
from $1$ to $M$.  An estimate of your uncertainty variance on $m$ is
\begin{equation}
\sigma_m^2 = \frac{1}{M}\,\sum_{j=1}^M [m_j-m]^2 \quad ,
\end{equation}
where $m$ stands for the best-fit $m$ using all the data.  The
uncertainty variance on $b$ is the same but with $[b_j-b]^2$ in the
sum, and the covariance $\sigma_{mb}$ is the same but with
$[m_j-m]\,[b_j-b]$.   Bootstrap
creates a new parameter, the number $M$ of trials.  There is a huge
literature on this, so we won't say anything too specific, but one
intuition is that once you have $M$ comparable to $N$, there probably
isn't much else you can learn, unless you got terribly unlucky with
your random number generator.

In jackknife, you make your measurement $N$ times, each time
\emph{leaving out} data point $i$.  Again, it doesn't matter what
method you are using, for each leave-one-out trial $i$ you get an
estimate $(m_i,b_i)$ found by fitting with all the data \emph{except}
point $i$.  Then you calculate
\begin{equation}
m = \frac{1}{N}\,\sum_{i=1}^N m_i \quad ,
\end{equation}
and the uncertainty variance becomes
\begin{equation}
\sigma_m^2 = \frac{N-1}{N}\,\sum_{i=1}^N [m_i-m]^2 \quad ,
\end{equation}
with the obvious modifications to make $\sigma_b^2$ and $\sigma_{mb}$.
The factor $[N-1]/N$ accounts, magically, for the fact that the
samples are not independent in any sense; this factor can only be
justified in the limit that everything is Gaussian and all the points
are identical in their noise properties.

Jackknife and bootstrap are both extremely useful when you don't know
or don't trust your uncertainties $\allsigmay$.  However, they
\emph{do} make assumptions about the problem.  For example, they can
only be justified when data represent a reasonable sampling from some
stationary frequency distribution of \emph{possible data} from a set
of hypothetical, similar experiments.  This assumption is, in some
ways, very strong, and often violated.\note{Neither jackknife nor
  bootstrap make sense, in their naive forms, if there is a
  significant dynamic range in the uncertainty variances
  $\sigma_{yi}^2$ of the data points.  This is because these
  techniques effectively treat all of the data equally.  Of course,
  both methods could be adjusted to account for this; bootstrap could
  be made to randomly select points somehow proportionally to their
  contribution to the total inverse uncertainty variance (it is the
  sum of the $\sigma_{yi}^{-2}$ that determines the total amount of
  information), and jackknife trials could be combined in a weighted
  way, weighted by the total inverse uncertainty variance.  But these
  generalizations require significant research; probably the methods
  should be avoided if there is a large diversity in the
  $\sigma_{yi}^2$.}  For another example, these methods will not give
correct answers if the model does not fit the data reasonably well.
That said, if you \emph{do} believe your uncertainty estimates, any
differences between the jackknife, bootstrap, and traditional
uncertainty estimates for the best-fit line parameters undermine
confidence in the validity of the model.  If you believe the model,
then any differences between the jackknife, bootstrap, and traditional
uncertainty estimates undermine confidence in the data-point
uncertainty variances $\allsigmay$.

\begin{problem}
Compute the standard uncertainty $\sigma_m^2$ obtained for the slope
of the line found by the standard fit you did in
\problemname~\ref{prob:standard}.  Now make jackknife (20 trials) and
bootstrap estimates for the uncertainty $\sigma_m^2$.  How do the
uncertainties compare and which seems most reasonable, given the data
and uncertainties on the data?
\end{problem}

\begin{problem}\label{prob:reduceerror}
Re-do \problemname~\ref{prob:mixture}---the mixture-based outlier
model---but just with the ``inlier'' points 5 through 20 from
\tablename~\ref{table:data_allerr} on
page~\pageref{table:data_allerr}.  Then do the same again, but with
all measurement uncertainties reduced by a factor of 2 (uncertainty
variances reduced by a factor of 4).  Plot the marginalized posterior
probability distributions for line parameters $(m,b)$ in both cases.
Your plots should look like those in
\figurename~\ref{fig:reduceerror}.  Did these posterior distributions
get smaller or larger with the reduction in the data-point
uncertainties?  Compare this with the dependence of the standard
uncertainty estimate $\inverse{\left[\mAT\,\mCinv\,\mA\right]}$.
\end{problem}

\section{Non-Gaussian uncertainties}\label{sec:non-Gaussian}

The standard method of \sectionname~\ref{sec:standard} is only
justified when the measurement uncertainties are Gaussian with known
variances.  Many noise processes in the real world are \emph{not}
Gaussian; what to do?  There are three general kinds of methodologies
for dealing with non-Gaussian uncertainties.  Before we describe them, permit
an aside on uncertainty and error.

In \notename~\ref{note:error}, we point out that an \emph{uncertainty} is
a measure of what you don't know, while an \emph{error} is a mistake
or incorrect value.  The fact that a measurement $y_i$ differs from
some kind of ``true'' value $Y_i$ that it would have in some kind of
``perfect experiment'' can be for two reasons (which are not
unrelated): It could be that there is some source of \emph{noise} in
the experiment that generates offsets for each data point $i$ away
from its true value.  Or it could be that the ``measurement'' $y_i$ is
actually the outcome of some probabilistic inference \emph{itself} and
therefore comes with a posterior probability distribution.  In either
case, there is a finite uncertainty, and in both cases it comes
somehow from imperfection in the experiment, but the way you
understand the distribution of the uncertainty is different.  In the
first case, you have to make a model for the noise in the experiment.
In the second, you have to analyze some posteriors.  In either case,
clearly, the uncertainty can be non-Gaussian.

All that said, and even when you \emph{suspect} that the sources of
noise are not Gaussian, it is still sometimes okay to treat the
\emph{uncertainties} as Gaussian: The Gaussian distribution is the
maximum-entropy distribution for a fixed \emph{variance}.  That means
that if you have some (perhaps magical) way to estimate the variance
of your uncertainty, and you aren't sure of anything else but the
variance, then the Gaussian is---contrary to popular opinion---the
\emph{most conservative} thing you can assume.\note{Assuming a
  Gaussian form is only conservative when you truly know the
  \emph{variance} of the noise.  Gaussians penalize heavily outliers,
  so they are not conservative in most situations.  But if you
  included the outliers in your variance calculation, then it is
  conservative to make the Gaussian assumption.  The issue is that it
  is almost impossible to make a variance estimate that captures the
  outliers properly.}  Of course it is very rare that you do know the
variance of your uncertainty or noise; most simple uncertainty
estimators are measurements of the central part of the distribution
function (not the tails) and are therefore \emph{not} good estimators
of the total variance.

Back to business.  The first approach to non-Gaussian noise or
uncertainty is (somehow) to estimate the true or total variance of
your uncertainties and then do the most conservative thing, which is
to once again assume that the noise is Gaussian!  This is the simplest
approach, but rarely possible, because when there are non-Gaussian
uncertainties, the observed variance of a finite sample of points can
be very much smaller than the true variance of the generating
distribution function.  The second approach is to make no attempt to
understand the non-Gaussianity, but to use the data-rejection methods
of \sectionname~\ref{sec:outliers}.  We generally recommend
data-rejection methods, but there is a third approach in which the
likelihood is artificially or heuristically softened to reduce the
influence of the outliers.  This approach is not usually a good idea
when it makes the objective function far from anything
justifiable.\note{We don't recommend them, but there are many
  non-Gaussian methods for removing sensitivity to outliers, which
  involve softening the objective function from $\chi^2$ (which
  depends quadratically on all residuals) to something that depends on
  large residuals to a smaller power.  The most straightforward way to
  soften the objective function is to lower the power to which
  residuals are raised.  For example, if we model the frequency
  distribution $p(y_i|x_i,\sigma_{yi},m,b)$ not with a Gaussian but
  rather with a biexponential
  \begin{equation}
  p(y_i|x_i,\sigma_{yi},m,b) = \frac{1}{2\,s_i}
   \,\exp\left(-\frac{|y_i-m\,x_i-b|}{s_i}\right) \quad ,
  \end{equation}
  where $s_i$ is an estimate of the mean absolute uncertainty,
  probably correctly set to something like $s_i\approx \sigma_{yi}$.
  Optimization of the total log likelihood is equivalent to minimizing
  \begin{equation}\label{eq:biexp}
  X = \sum_{i=1}^N \frac{|y_i-f(x_i)|}{s_i} \quad ,
  \end{equation}
  where $f(x)$ is the straight line of \equationname~\ref{eq:fofx}.
  This approach is rarely justified, but it has the nice
  property that it introduces no new parameters.
  Another straightforward softening is to (smoothly) cut off the
  contribution of a residual as it becomes large.  For example, replace
  $\chi^2$ with
  \begin{equation}\label{eq:soft}
  \chi_Q^2 = \sum_{i=1}^N \frac{Q^2\,[y_i-f(x_i)]^2}
    {Q^2\,\sigma_{yi}^2+[y_i-f(x_i)]^2} \quad ,
  \end{equation}
  where $Q^2$ is the maximum amount a point can contribute to
  $\chi_Q^2$ \citep{hampel}.  When each residual is small, its
  contribution to $\chi_Q^2$ is nearly identical to its contribution
  to the standard $\chi^2$, but as the residual gets substantial, the
  contribution of the residual does not increase as the square.  This
  is about the simplest robust method that introduces only one new
  parameter ($Q$), and in principle, one can put a prior on $Q$, infer
  it, and marginalize it out.  In general these are both far less good
  alternatives than modeling the outliers, because objective functions
  softer than $\chi^2$ are rarely justifiable in terms of any
  generative model for the data.  In this \sectionname\ we modeled
  outliers by a method that can almost be interpreted in terms of a
  softer $\chi^2$; in \sectionname~\ref{sec:non-Gaussian} we discuss
  (briefly) justifiable modifications to the objective function when
  the observational uncertainty distributions depart from Gaussians in
  known ways.  Nonetheless, even when there is not a good
  justification for softening the objective function, it can sometimes
  be useful for generating a robust fit with a minimum of effort.

  Along the same lines, sometimes it \emph{is} justifiable to use
  Student's $t$-distribution for a softened objective.  That is
  because the $t$-distribution is what you get when you believe that
  the uncertainties are Gaussian, you don't know their variances, but
  you imagine that the variances are drawn from some particular
  distribution (inverse chi-squared), and you marginalize.}  The
fourth approach---the only approach that is really justified---is to
fully understand and model the non-Gaussianity.

For example, if data points are generated by a process the noise from
which looks like a sum or mixture of $k$ Gaussians with different
variances, and the distribution can be understood well enough to be
modeled, the investigator can replace the Gaussian objective function
with something more realistic.  Indeed, \emph{any} (reasonable)
frequency distribution can be described as a mixture of Gaussians, so
this approach is extremely general (in principle; it may not be easy).
For example, the generative model in \sectionname~\ref{sec:objective}
for the standard case was built out of individual data-point
likelihoods given in \equationname~(\ref{eq:objectivei}).  If the
frequency distribution for the noise contribution to data point $y_i$
has been expressed as the sum of $k$ Gaussians, these likelihoods
become
\begin{equation}
p(y_i|x_i,\sigma_{yi},m,b) = \sum_{j=1}^k
 \frac{a_{ij}}{\sqrt{2\,\pi\,\sigma_{yij}^2}}
 \,\exp\left(-\frac{[y_i+\Delta y_{ij}-m\,x_i-b]^2}{2\,\sigma_{yij}^2}\right)
 \quad ,
\end{equation}
where the $k$ Gaussians for measurement $i$ have variances
$\sigma_{yij}$, offsets (means---there is no need for the Gaussians to
be concentric) $\Delta y_{ij}$, and amplitudes $a_{ij}$ that sum to
unity
\begin{equation}
\sum_{j=1}^k a_{ij} = 1
\end{equation}
In this case, fitting becomes much more challenging (from an
optimization standpoint), but this can become a \emph{completely
  general} objective function for arbitrarily complicated non-Gaussian
uncertainties.  It can even be generalized to situations of
arbitrarily complicated joint distributions for \emph{all} the
measurement uncertainties.

One very common non-Gaussian situation is one in which the data points
include upper or lower limits, or the investigator has a value with an
uncertainty estimate but knows that the true value can't enter into
some region (for example, there are many situations in which one knows
that all the true $Y_i$ must be greater than zero).  In all these
situations---of upper limits or lower limits or otherwise limited
uncertainties---the best thing to do is to model the uncertainty
distribution as well as possible, construct the proper justified
scalar objective function, and carry on.\note{The possibility arises
  of having data points that are not true measurements but only, for
  example, upper limits.  In most contexts (in astrophysics, anyway)
  an ``upper limit'' is caused by one of two things.  Either the
  investigator has taken the logarithm of a quantity that in fact, for
  some measurements, went to zero or negative; or else the
  investigator is fitting a model to a binned quantity and in some
  bins there are no objects or data.  In the first case, the best
  advice is not to take the logarithm at all.  Fit a power-law or
  exponential in the linear space rather than a straight line in the
  log space.  In the second case, the best advice is to use the
  likelihood appropriate for a Poisson process rather than a Gaussian
  process.  Least-square fitting is an approximation when the data are
  generated by a Poisson sampling; this approximation breaks down when
  the number of objects per bin becomes small.  Optimizing the correct
  Poisson likelihood is not difficult, and it is the only
  scientifically justifiable thing to do.

  It is rare that there are true upper limits not caused by one of the
  above.  However, if neither of the above applies---perhaps because
  the investigator was given only the limits, and not the details of
  their origins---the right thing to do is to write down some
  generative model for the limits.  This would be some description of
  the ``probability of the data given the model'' where the (say)
  95-percent upper limit is replaced with a probability distribution
  that has 95~percent of its mass below the reported limit value.
  This is ideally done with a correct model of the measurement
  process, but it is acceptable in cases of justifiable ignorance to
  make up something reasonable; preferably something that is close to
  whatever distribution maximizes the entropy given the reported and
  logical constraints.}  Optimization might become challenging, but
that is an engineering problem that must be tackled for scientific
correctness.

\section{Goodness of fit and unknown uncertainties}\label{sec:goodness}

How can you decide if your fit is a good one, or that your assumptions
are justified?  And how can you infer or assess the individual
data-point uncertainties if you don't know them, or don't trust them?
These seem like different questions, but they are coupled: In the
standard (frequentist) paradigm, you can't test your assumptions
unless you are very confident about your uncertainty estimates, and
you can't test your uncertainty estimates unless you are very
confident about your assumptions.

Imagine that the generative model in \sectionname~\ref{sec:objective}
is a valid description of the data.  In this case, the noise
contribution for each data point $i$ has been drawn from a Gaussian
with variance $\sigma_{yi}^2$.  The expectation is that data point $i$
will provide a mean squared error comparable to $\sigma_{yi}^2$, and a
contribution of order unity to $\chi^2$ when the parameters $(m,b)$
are set close to their true values.  Because in detail there are two
fit parameters $(m,b)$ which you have been permitted to optimize, the
expectation for $\chi^2$ is smaller than $N$. The model is linear, so
the distribution of $\chi^2$ is known analytically and is given
(unsurprisingly) by a \emph{chi-square} distribution: in the limit of
large $N$ the rule of thumb is that---when the model is a good fit,
the uncertainties are Gaussian with known variances, and there are two
linear parameters ($m$ and $b$ in this case),
\begin{equation}
\chi^2 = [N-2] \pm \sqrt{2\,[N-2]} \quad ,
\end{equation}
where the $\pm$ symbol is used loosely to indicate something close to
a standard uncertainty (something close to a 68-percent confidence
interval).  If you find $\chi^2$ in this ballpark, it is conceivable
that the model is good.\note{Frequentist model rejection on the basis
  of unacceptable $\chi^2$ should not be confused with model
  \emph{comparison} performed with $\chi^2$.  If $\chi^2$ differs from
  the number of degrees of freedom $[N-2]$ by something on the order
  of $\sqrt{2\,[N-2]}$, the model cannot be rejected on a $\chi^2$
  basis.  However, if two equally plausible models differ from one
  another by a difference $\Delta\chi^2$ substantially larger than
  unity, the model with the lower $\chi^2$ can be preferred and the
  higher rejected on that basis, even if the higher still shows an
  acceptable $\chi^2$.  This difference between model rejection and
  model comparison is counterintuitive, but it emerges from the
  difference that in the model comparison case, there are \emph{two}
  models, not one, and they are both (by assumption) equally plausible
  \apriori.  The conditions for model comparison are fairly strong, so
  the test is fairly sensitive.  In the case of correctly estimated
  uncertainty variances and two models that are equally plausible
  \apriori, the $\Delta\chi^2$ value between the models is (twice) the
  natural logarithm of the odds ratio between the models.  For this
  reason, large $\Delta\chi^2$ values generate considerable
  confidence.}

Model rejection on the basis of too-large $\chi^2$ is a frequentist's
option.  A Bayesian can't interpret the data without a model, so there
is no meaning to the question ``is this model good?''.  Bayesians only
answer questions of the form ``is this model better than that one?''.
Because we are only considering the straight-line model in this
\documentname, further discussion of this (controversial, it turns
out) point goes beyond our scope.\note{Technically, Bayesians cannot
  reject a model outright on the basis of bad $\chi^2$ alone;
  Bayesians can only \emph{compare} comparable models.  A Bayesian
  will only consider the absolute value of $\chi^2$ to provide hints
  about what might be going wrong.  Part of the reason for this is
  that a Bayesian analysis returns a probability distribution over
  mutually exclusive hypotheses.  This distribution or this set of
  probabilities integrates or sums to unity.  If only one model (``the
  data are drawn from a straight line, plus noise'') is considered,
  the probabilities for the parameters of that model will integrate to
  unity and the model cannot be disfavored in any sense.  Of course,
  any time an investigator considers only one model, he or she is
  almost certainly making a mistake, and any conservative Bayesian
  data analyst will consider a large space of models and ask about the
  evidence for the alternatives, or for complexification of the
  current model.  So a Bayesian \emph{can} reject models on a $\chi^2$
  basis---provided that he or she has permitted a wide enough range of
  models, and provided those models have somewhat comparable prior
  probabilities.  Model rejection and hypothesis comparison are large
  topics that go beyond the scope of this \documentname.

  Many Bayesians argue that they \emph{can} reject models without
  alternatives, by, for example, comparing likelihood (or Bayes
  factor---the likelihood integrated over the prior) under the data
  and under artificial data generated with the same model (with
  parameters drawn from the prior).  This is true, and it is a good
  idea for any Bayesian to perform this test, but there is still no
  correct proababilistic statement that can be made about model
  rejection in the single-model situation (you can't reject your only
  model).

  In general, the investigator's decision-making about what models to
  consider (which will be informed by tests like comparisons between
  likelihoods and fake-data likelihoods) have an enormous impact on
  her or his scientific conclusions.  At the beginning of an
  investigation, this problem lies in the setting of priors.  An
  investigator who considers \emph{only} a straight-line model for her
  or his points is effectively setting insanely informative
  priors---the prior probability of every alternative model is
  precisely zero!  At the end of an investigation, this problem---what
  models to report or treat as confirmed---lies in the area of
  \emph{decision theory}, another topic beyond the scope of this
  \documentname.  But even in a context as limited as straight-line
  fitting, it is worth remembering that, as scientists, we are not
  just inferring things, we are making \emph{decisions}---about what
  is right and wrong, about what to investigate, about what to report,
  and about what to advocate---and these decisions are only indirectly
  related to the inferences we perform.}

It is easy to get a $\chi^2$ much \emph{higher} than $[N-2]$; getting
a \emph{lower} value seems impossible; yet it happens very frequently.
This is one of the many reasons that rejection or acceptance of a
model on the $\chi^2$ basis is dangerous; exactly the same kinds of
problems that can make $\chi^2$ unreasonably low can make it
unreasonably high; worse yet, it can make $\chi^2$ reasonable when the
model is bad.  Reasons the $\chi^2$ value can be lower include that
uncertainty estimates can easily be overestimated. The opposite of
this problem can make $\chi^2$ high when the model is good. Even worse
is the fact that uncertainty estimates are often correlated, which can
result in a $\chi^2$ value significantly different from the expected
value.

Correlated measurements are not uncommon; the process by which the
observations were taken often involve shared information or
assumptions. If the individual data-points $y_i$ have been estimated
by some means that effectively relies on that shared information, then
there will be large covariances among the data points.  These
covariances bring off-diagonal elements into the covariance matrix
$\mC$, which was trivially constructed in
\equationname~(\ref{eq:covar}) under the assumption that all
covariances (off-diagonal elements) are precisely zero.  Once the
off-diagonal elements are non-zero, $\chi^2$ must be computed by the
matrix expression in \equationname~(\ref{eq:chisquared}); this is
equivalent to replacing the sum over $i$ to two sums over $i$ and $j$
and considering all cross terms
\begin{equation}
\chi^2 =
 \transpose{\left[\mY-\mA\,\mX\right]}\,\mCinv\,\left[\mY-\mA\,\mX\right]
 = \sum_{i=1}^N \sum_{j=1}^N
 w_{ij}\,\left[y_i-f(x_i)\right]\,\left[y_j-f(x_j)\right]
 \quad,
\end{equation} 
where the $w_{ij}$ are the elements of the inverse covariance
matrix $\inverse{\mC}$.

In principle, data-point uncertainty variance underestimation or mean
point-to-point covariance can be estimated by adjusting them until
$\chi^2$ is reasonable, in a model you know (for independent reasons)
to be good.  This is rarely a good idea, both because you rarely know
that the model is a good one, and because you are much better served
understanding your variances and covariances directly.

If you don't have or trust your uncertainty estimates and don't care
about them at all, your best bet is to go Bayesian, infer them, and
marginalize them out.  Imagine that you don't know anything about your
individual data-point uncertainty variances $\allsigmay$ at all.
Imagine, further, that you don't even know anything about their
\emph{relative} magnitudes; that is, you can't even assume that they
have similar magnitudes.  In this case a procedure is the following:
Move the uncertainties into the model parameters to get the large
parameter list $(m,b,\allsigmay)$.  Pick a prior on the uncertainties
(on which you must have \emph{some} prior information, given, for
example, the range of your data and the like).  Apply Bayes's rule to
obtain a posterior distribution function for all the parameters
$(m,b,\allsigmay)$.  Marginalize over the uncertainties to obtain the
properly marginalized posterior distribution function for $(m,b)$.  No
sane person would imagine that the procedure described here can lead
to any informative inference.  However, if the prior on the
$\allsigmay$ is relatively flat in the relevant region, the \emph{lack
  of specificity} of the model when the uncertainties are large pushes
the system to smaller uncertainties, and the \emph{inaccuracy} of the
model when the uncertainties are small pushes the system to larger
uncertainties.  If the model is reasonable, the inferred uncertainties
will be reasonable too.

\begin{problem}
Assess the $\chi^2$ value for the fit performed in
\problemname~\ref{prob:easy} (do that problem first if you haven't
already).  Is the fit good?  What about for the fit performed in
\problemname~\ref{prob:standard}?
\end{problem}

\begin{figure}[htbp]
\exampleplottwo{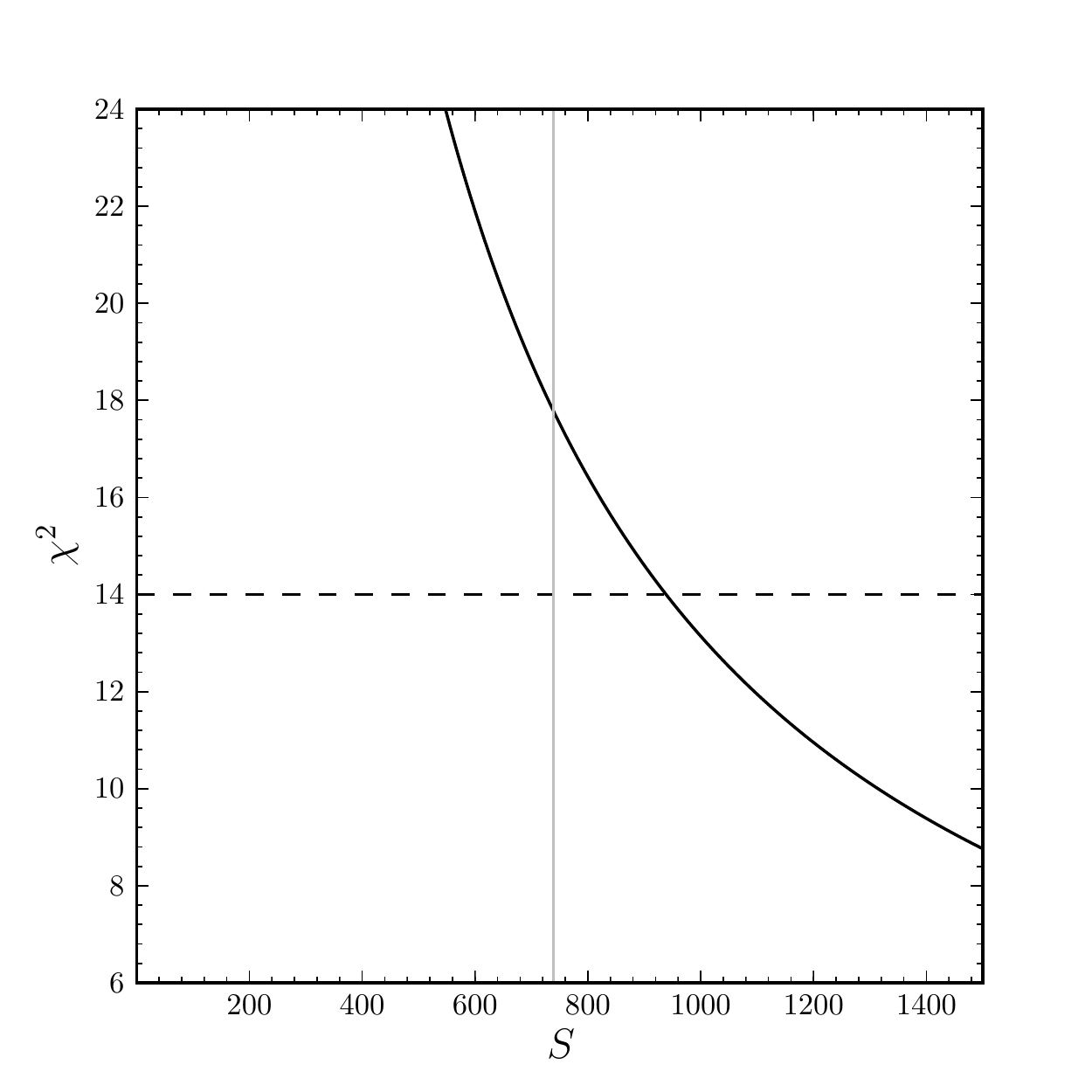}{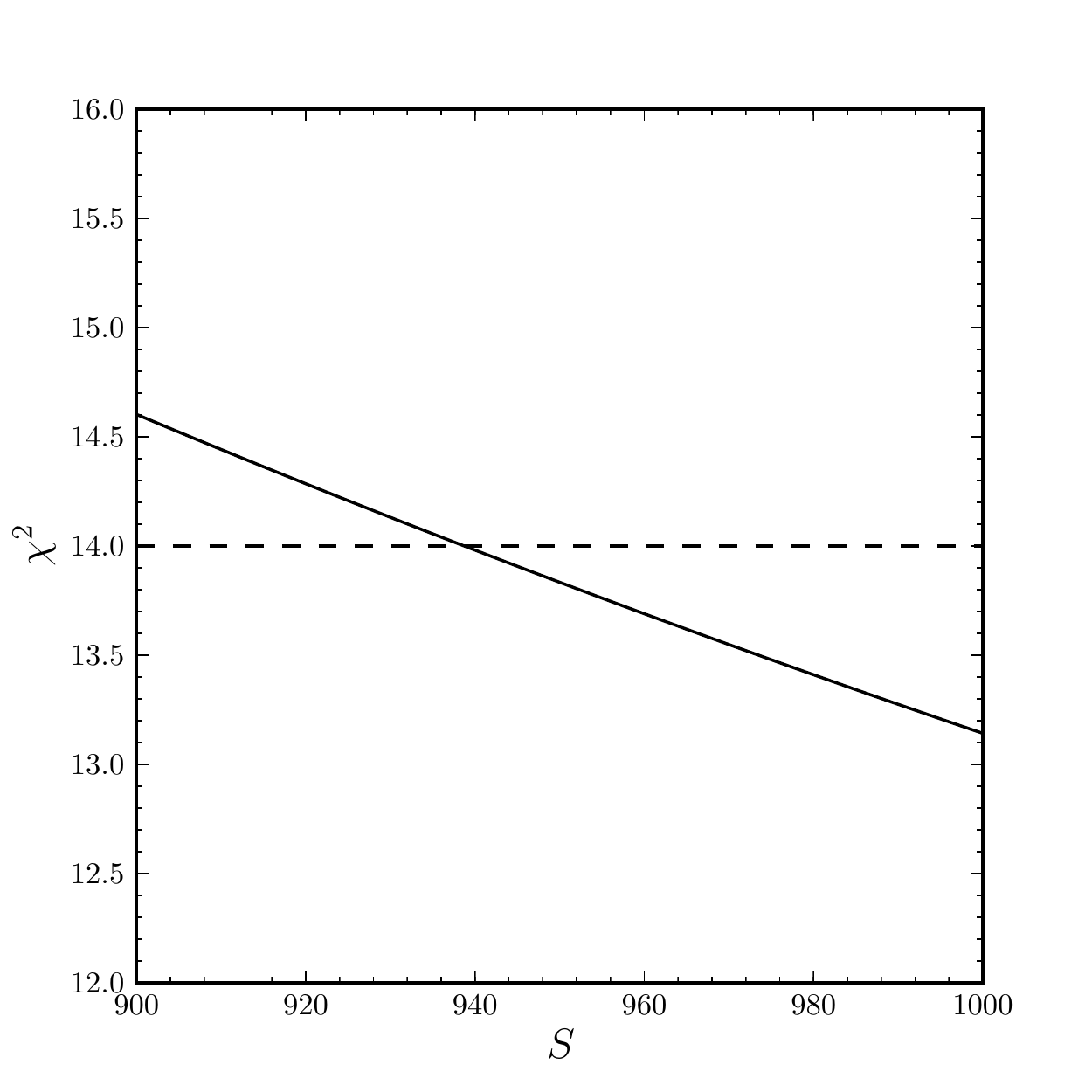}
\caption{Partial solution to \problemname~\ref{prob:chi2}: The
dependence of the fitting scalar $\chi^2$ on what is assumed about the
data uncertainties.}\label{fig:chi2}
\end{figure}

\begin{problem}\label{prob:chi2}
Re-do the fit of \problemname~\ref{prob:easy} but setting all
$\sigma_{yi}^2=S$, that is, ignoring the uncertainties and replacing
them all with the same value $S$.  What uncertainty variance $S$ would
make $\chi^2 = N-2$?  Relevant plots are shown in
\figurename~\ref{fig:chi2}.  How does it compare to the mean and
median of the uncertainty variances $\allsigmay$?
\end{problem}

\begin{figure}[htbp]
\exampleplottwo{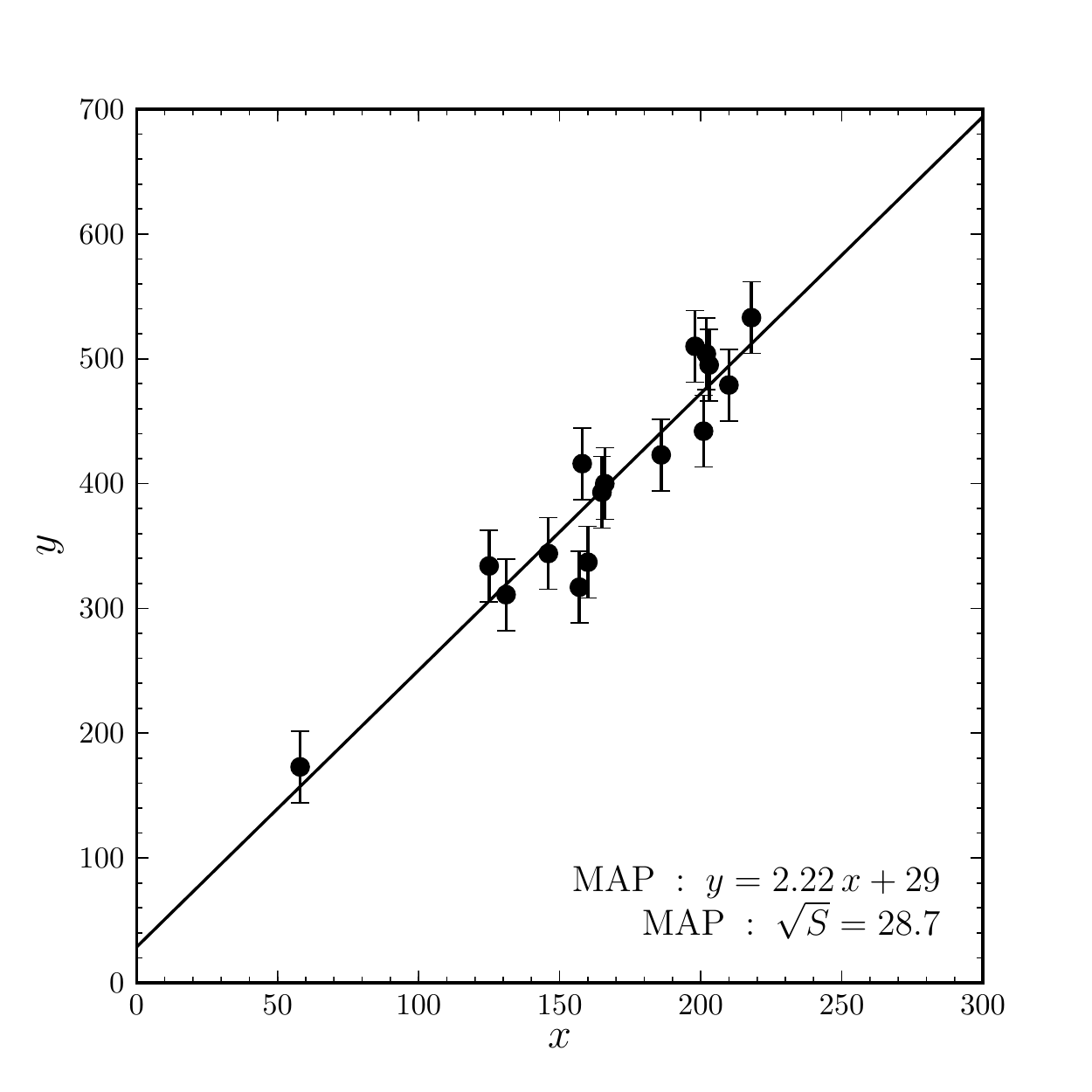}{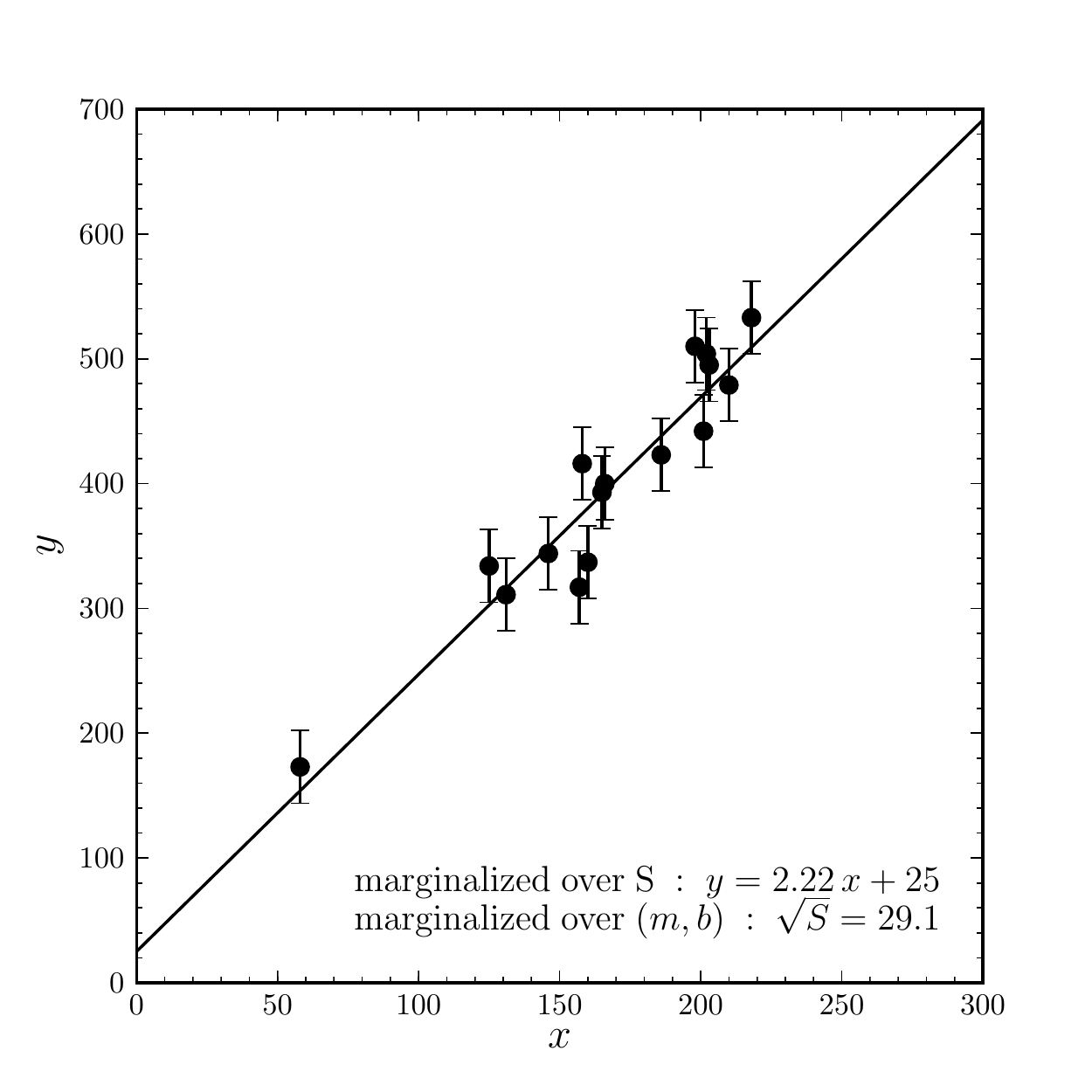}
\caption{Partial solution to \problemname~\ref{prob:bayeschi2}: The
results of inferring the observational uncertainties simultaneously
with the fit parameters.}\label{fig:bayeschi2}
\end{figure}

\begin{problem}\label{prob:bayeschi2}
Flesh out and write all equations for the Bayesian uncertainty
estimation and marginalization procedure described in this
\sectionname.  Note that the inference and marginalization would be
very expensive without excellent sampling tools!  Make the additional
(unjustified) assumption that all the uncertainties have the same
variance $\sigma_{yi}^2=S$ to make the problem tractable.  Apply the
method to the $x$ and $y$ values for points 5 through 20 in
\tablename~\ref{table:data_allerr} on
page~\pageref{table:data_allerr}.  Make a plot showing the points, the
maximum \aposteriori\ value of the uncertainty variance as error bars,
and the maximum \aposteriori\ straight line.  For extra credit, plot
two straight lines, one that is maximum \aposteriori\ for the full
posterior and one that is the same but for the posterior after the
uncertainty variance $S$ has been marginalized out.  Your result
should look like \figurename~\ref{fig:bayeschi2}.  Also plot two
sets of error bars, one that shows the maximum for the full posterior
and one for the posterior after the line parameters $(m,b)$ have been
marginalized out.
\end{problem}

\section{Arbitrary two-dimensional uncertainties}\label{sec:twod}

Of course most real two-dimensional data $\allxy$ come with
uncertainties in \emph{both} directions (in both $x$ and $y$).  You
might not know the amplitudes of these uncertainties, but it is
unlikely that the $x$ values are known to sufficiently high accuracy
that any of the straight-line fitting methods given so far is valid.
Recall that everything so far has assumed that the $x$-direction
uncertainties were negligible.  This might be true, for example, when
the $x_i$ are the times at which a set of stellar measurements are
made, and the $y_i$ are the declinations of the star at those times.
It is \emph{not} going to be true when the $x_i$ are the right
ascensions of the star.

In general, when one makes a two-dimensional measurement $(x_i,y_i)$,
that measurement comes with uncertainties $(\sigma_{xi}^2,\sigma_{yi}^2)$
in both directions, and some covariance $\sigma_{xyi}$ between them.
These can be put together into a covariance tensor $\mS_i$
\begin{equation}
\mS_i \equiv \left[\begin{array}{cc}
\sigma_{xi}^2 & \sigma_{xyi} \\ \sigma_{xyi} & \sigma_{yi}^2
\end{array}\right] \quad .
\end{equation}
If the uncertainties are Gaussian, or if all that is known about the
uncertainties is their variances, then the covariance tensor can be
used in a two-dimensional Gaussian representation of the probability
of getting measurement $(x_i,y_i)$ when the ``true value'' (the value
you would have for this data point if it had been observed with
negligible noise) is $(x,y)$:
\begin{equation}
p(x_i,y_i|\mS_i,x,y) = \frac{1}{2\,\pi\,\sqrt{\det(\mS_i)}}
  \,\exp\left(-\frac{1}{2}\,\transpose{\left[\mZ_i - \mZ\right]}
  \,\inverse{\mS_i}\,\left[\mZ_i - \mZ\right]\right) \quad ,
\end{equation}
where we have implicitly made column vectors
\begin{equation}\label{eq:mZ}
\mZ = \left[\begin{array}{c} x \\ y \end{array}\right] \quad ; \quad
\mZ_i = \left[\begin{array}{c} x_i \\ y_i \end{array}\right] \quad .
\end{equation}

Now in the face of these general (though Gaussian) two-dimensional
uncertainties, how do we fit a line?  Justified objective functions
will have something to do with the probability of the observations
$\allxy$ given the uncertainties $\allS$, as a function of properties
$(m,b)$ of the line.  As in \sectionname~\ref{sec:objective}, the
probability of the observations given model parameters $(m,b)$ is
proportional to the \emph{likelihood} of the parameters given the
observations.  We will now construct this likelihood and maximize it,
or else multiply it by a prior on the parameters and report the
posterior on the parameters.

Schematically, the construction of the likelihood involves specifying
the line (parameterized by $m,b$), finding the probability of each
observed data point given any true point on that line, and
marginalizing over all possible true points on the line.  This is a
model in which each point really does have a true location on the
line, but that location is not directly measured; it is, in some
sense, a missing datum for each point.\note{One amusing aspect of
  the generative model we are about to write down is that it is
  possible to marginalize it over \emph{all} the parameters and all
  the true positions of the points along the linear relationship
  except for one and thereby produce a marginalized posterior
  probability distribution function for the true position of any
  individual data point, under the assumptions of the generative
  model.  That is, it is possible to reconstruct the missing data.}
One approach is to think of the straight line as a two-dimensional
Gaussian with an infinite eigenvalue corresponding to the direction of
the line and a zero eigenvalue for the direction orthogonal to
this. In the end, this line of argument leads to \emph{projections} of
the two-dimensional uncertainty Gaussians along the line (or onto the
subspace that is orthogonal to the line), and evaluation of those at
the \emph{projected displacements}.  Projection is a standard linear
algebra technique, so we will use linear algebra (matrix)
notation.\note{The fact that the marginalization over the true
  positions of the points reduces to a projection along the line---or
  onto the subspace orthogonal to the line---makes this method very
  similar to that called ``orthogonal least squares''.  This
  generative model justifies that procedure, in the case when the
  orthogonal displacements are penalized by the properly projected
  uncertainties.}

A slope $m$ can be described by a unit vector $\vhat$
\emph{orthogonal} to the line or linear relation:
\begin{equation}
\vhat
 = \frac{1}{\sqrt{1+m^2}}\,\left[\begin{array}{c}-m\\1\end{array}\right]
 = \left[\begin{array}{c}-\sin\theta\\\cos\theta\end{array}\right] \quad ,
\end{equation}
where we have defined the angle $\theta = \arctan m$ made between the
line and the $x$ axis.  The orthogonal displacement $\Delta_i$ of each
data point $(x_i,y_i)$ from the line is given by
\begin{equation}
\Delta_i = \transpose{\vhat}\,\mZ_i - b\,\cos\theta \quad ,
\end{equation}
where $\mZ_i$ is the column vector made from $(x_i,y_i)$ in
\equationname~(\ref{eq:mZ}).  Similarly, each data point's covariance
matrix $\mS_i$ projects down to an orthogonal variance $\Sigma_i^2$ given by
\begin{equation}\label{eq:Sigma}
\Sigma_i^2 = \transpose{\vhat}\,\mS_i\,\vhat
\end{equation}
and then the log likelihood for $(m,b)$ or $(\theta,b\,\cos\theta)$
can be written as
\begin{equation}\label{eq:twodlike}
\ln\like = K - \sum_{i=1}^N \frac{\Delta_i^2}{2\,\Sigma_{i}^2} \quad ,
\end{equation}
where $K$ is some constant.  \emph{This} likelihood can be maximized,
and the resulting values of $(m,b)$ are justifiably the best-fit
values.  The only modification we would suggest is performing the fit
or likelihood maximization not in terms of $(m,b)$ but rather
$(\theta,\bperp)$, where $\bperp\equiv[b\,\cos\theta]$ is the
perpendicular distance of the line from the origin.  This removes the
paradox that the standard ``prior'' assumption of standard
straight-line fitting treats all \emph{slopes} $m$ equally, putting
way too much attention on angles near $\pm\pi/2$.  The Bayesian must
set a prior.  Again, there are many choices, but the most natural
is \emph{something} like flat in $\theta$ and flat in $\bperp$ (the
latter not proper).

The implicit generative model here is that there are $N$ points with
true values that lie precisely on a narrow, linear relation in the
$x$--$y$ plane.  To each of these true points a Gaussian offset has
been added to make each observed point $(x_i,y_i)$, where the offset
was drawn from the two-dimensional Gaussian with covariance tensor
$\mS_i$.  As usual, if this generative model is a good approximation
to the properties of the data set, the method works very well.  Of
course there are many situations in which this is \emph{not} a good
approximation.  In \sectionname~\ref{sec:scatter}, we consider the
(very common) case that the relationship is near-linear but not
\emph{narrow}, so there is an intrinsic width or scatter in the true
relationship.  Another case is that there are outliers; this can be
taken care of by methods very precisely analogous to the methods in
\sectionname~\ref{sec:outliers}.  We ask for this in the
\problemnames\ below.

It is true that standard least-squares fitting is easy and simple;
presumably this explains why it is used so often when it is
inappropriate.  We hope to have convinced some readers that doing
something justifiable and sensible when there are uncertainties in
both dimensions---when standard linear fitting is inappropriate---is
neither difficult nor complex.  That said, there is something
fundamental wrong with the generative model of this \sectionname, and
it is that the model generates \emph{only} the displacements of the
points \emph{orthogonal} to the linear relationship.  The model is
completely unspecified for the distribution \emph{along} the
relationship.\note{This probably means that
  although this two-dimensional fitting method works, there is
  something fishy or improper involved.  The model is only strictly
  correct, in some sense, when the distribution \emph{along} the line
  is uniform over the entire range of interest.  That's rarely true in
  practice.\label{note:orthogonal}}

In the astrophysics literature (see, for example, the Tully--Fisher
literature), there is a tradition, when there are uncertainties in both
directions, of fitting the ``forward'' and ``reverse''
relations---that is, fitting $y$ as a function of $x$ and then $x$ as
a function of $y$---and then splitting the difference between the two
slopes so obtained, or treating the difference between the slopes as a
systematic uncertainty.  This is unjustified.\note{The difference
  between the forward-fitting and reverse-fitting slopes \emph{is} a
  systematic uncertainty in the sense that if you are doing something
  \emph{unjustifiable} you will certainly introduce large systematics!
  This forward--reverse procedure is not justifiable and it gives
  results which differ by substantially more than the uncertainty in
  the procedure given in this \sectionname.  We have already
  (\notename~\ref{note:regression}) criticized one of the
  principal papers to promote this kind of hack, and do not need to
  repeat that criticism here.}  Another common method for finding the
linear relationship in data when there are uncertainties in both directions
is \emph{principal components analysis}.  The manifold reasons
\emph{not} to use PCA are beyond the scope of this
\documentname.\note{Why not use principal components analysis?  In a
  nutshell: The method of PCA \emph{does} return a linear relationship
  for a data set, in the form of the dominant principal component.
  However, this is the dominant principal component of the observed
  data, not of the underlying linear relationship that, when noise is
  added, generates the observations.  For this reason, the output of
  PCA will be strongly drawn or affected by the individual data point
  noise covariances $\mS_i$.  This point is a subtle one, but in
  astrophysics it is almost certainly having a large effect in the
  standard literature on the ``fundamental plane'' of elliptical
  galaxies, and in other areas where a fit is being made to data with
  substantial uncertainties.  Of course PCA is another trivial, linear
  method, so it is often useful despite its inapplicability; and it
  becomes applicable in the limit that the data uncertainties are
  negligible (but in that case everything is trivial anyway).}

\begin{figure}[htbp]
\exampleplot{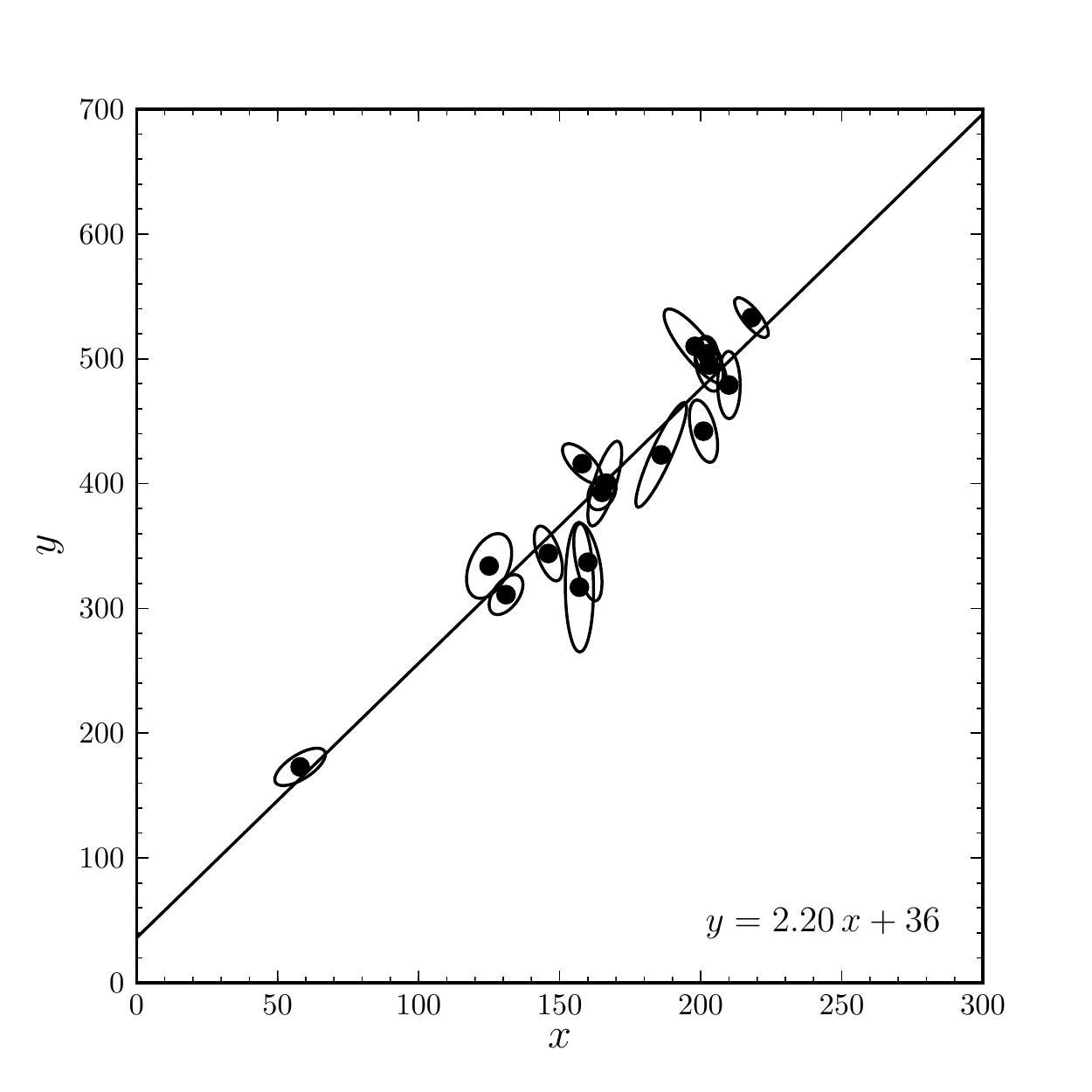}
\caption{Partial solution to \problemname~\ref{prob:twod}: The best
fit after properly accounting for covariant noise in both
dimensions.}\label{fig:twod}
\end{figure}

\begin{problem}\label{prob:twod}
Using the method of this \sectionname, fit the straight line
$y=m\,x+b$ to the $x$, $y$, $\sigma_x^2$, $\sigma_{xy}$, and
$\sigma_y^2$ values of points 5 through 20 taken from
\tablename~\ref{table:data_allerr} on
page~\pageref{table:data_allerr}.  Make a plot showing the points,
their two-dimensional uncertainties (show them as one-sigma ellipses),
and the best-fit line.  Your plot should look like
\figurename~\ref{fig:twod}.
\end{problem}

\begin{figure}[htbp]
\exampleplottwo{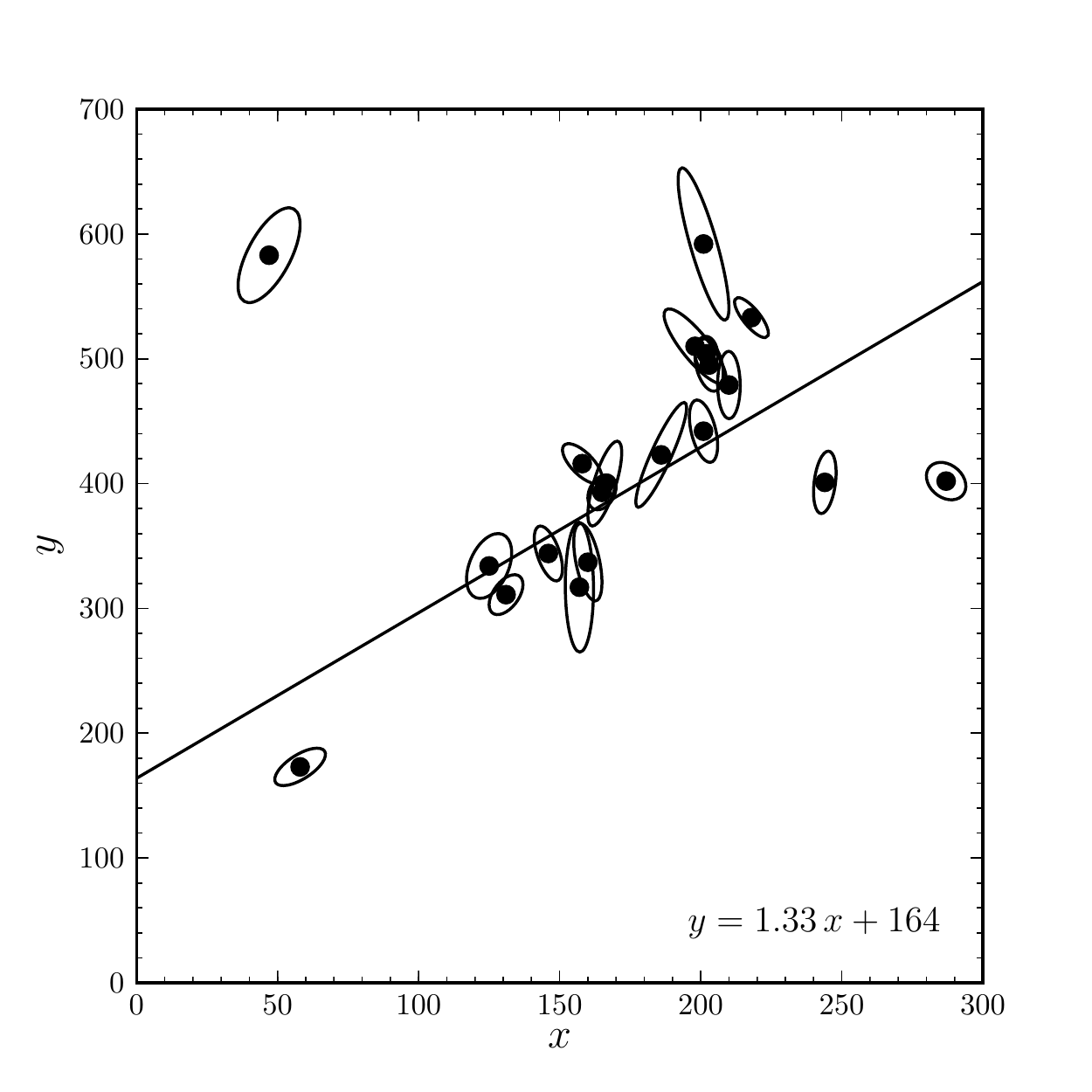}{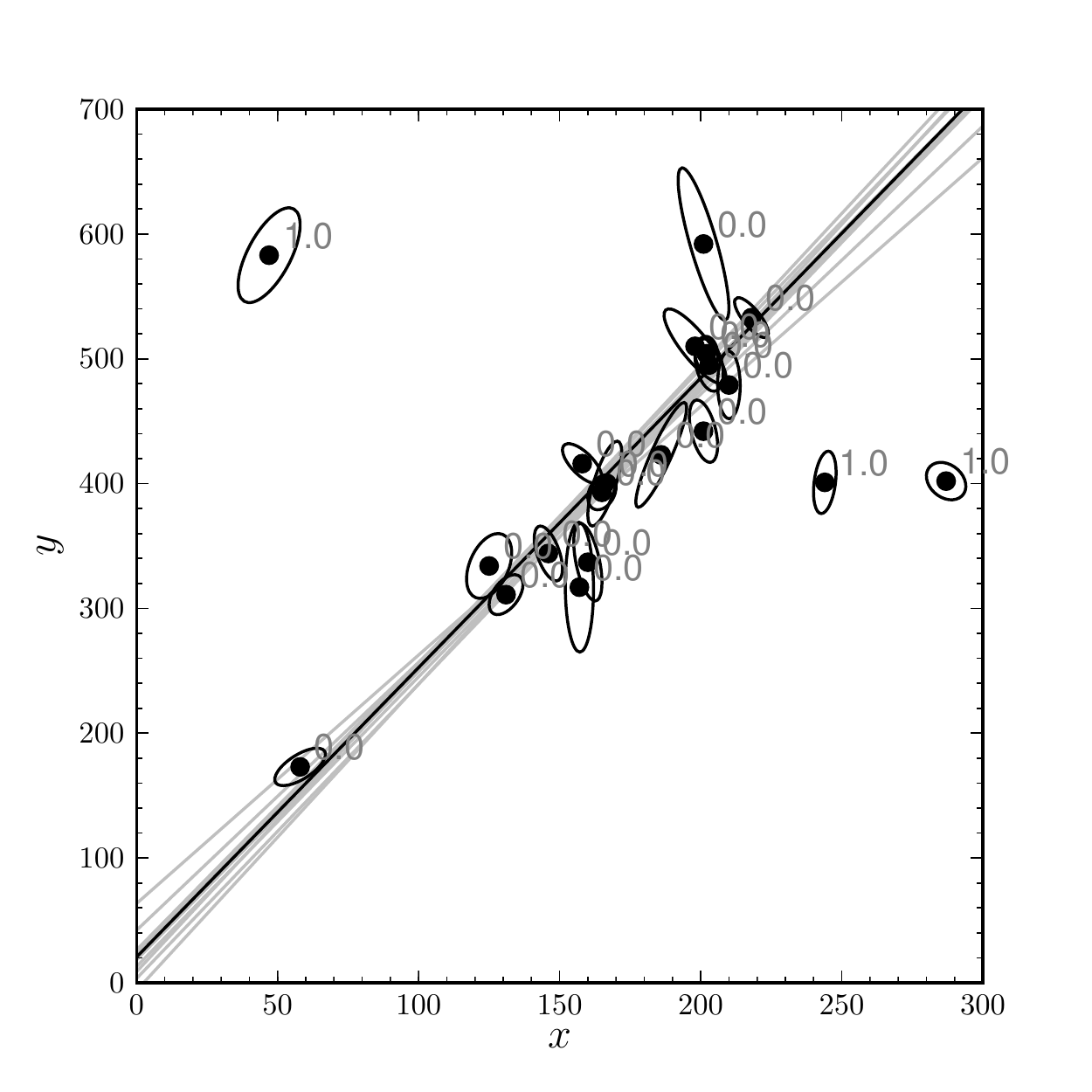}
\caption{Partial solution to \problemname~\ref{prob:twodoutlier}: On
the left, the same as \figurename~\ref{fig:twod} but including the
outlier points.  On the right, the same as in
\figurename~\ref{fig:mixture} but applying the outlier (mixture) model
to the case of two-dimensional uncertainties.}\label{fig:twodoutlier}
\end{figure}

\begin{problem}\label{prob:twodoutlier}
Repeat \problemname~\ref{prob:twod}, but using all of the data in
\tablename~\ref{table:data_allerr} on
page~\pageref{table:data_allerr}.  Some of the points are now
outliers, so your fit may look worse.  Follow the fit by a robust
procedure analogous to the Bayesian mixture model with bad-data
probability $\Pbad$ described in \sectionname~\ref{sec:outliers}.  Use
something sensible for the prior probability distribution for $(m,b)$.
Plot the two results with the data and uncertainties.  For extra
credit, plot a sampling of 10 lines drawn from the marginalized
posterior distribution for $(m,b)$ and plot the samples as a set of
light grey or transparent lines.  For extra extra credit, mark each
data point on your plot with the fully marginalized probability that
the point is bad (that is, rejected, or has $q=0$).  Your result
should look like \figurename~\ref{fig:twodoutlier}.
\end{problem}

\begin{figure}[htbp]
\exampleplot{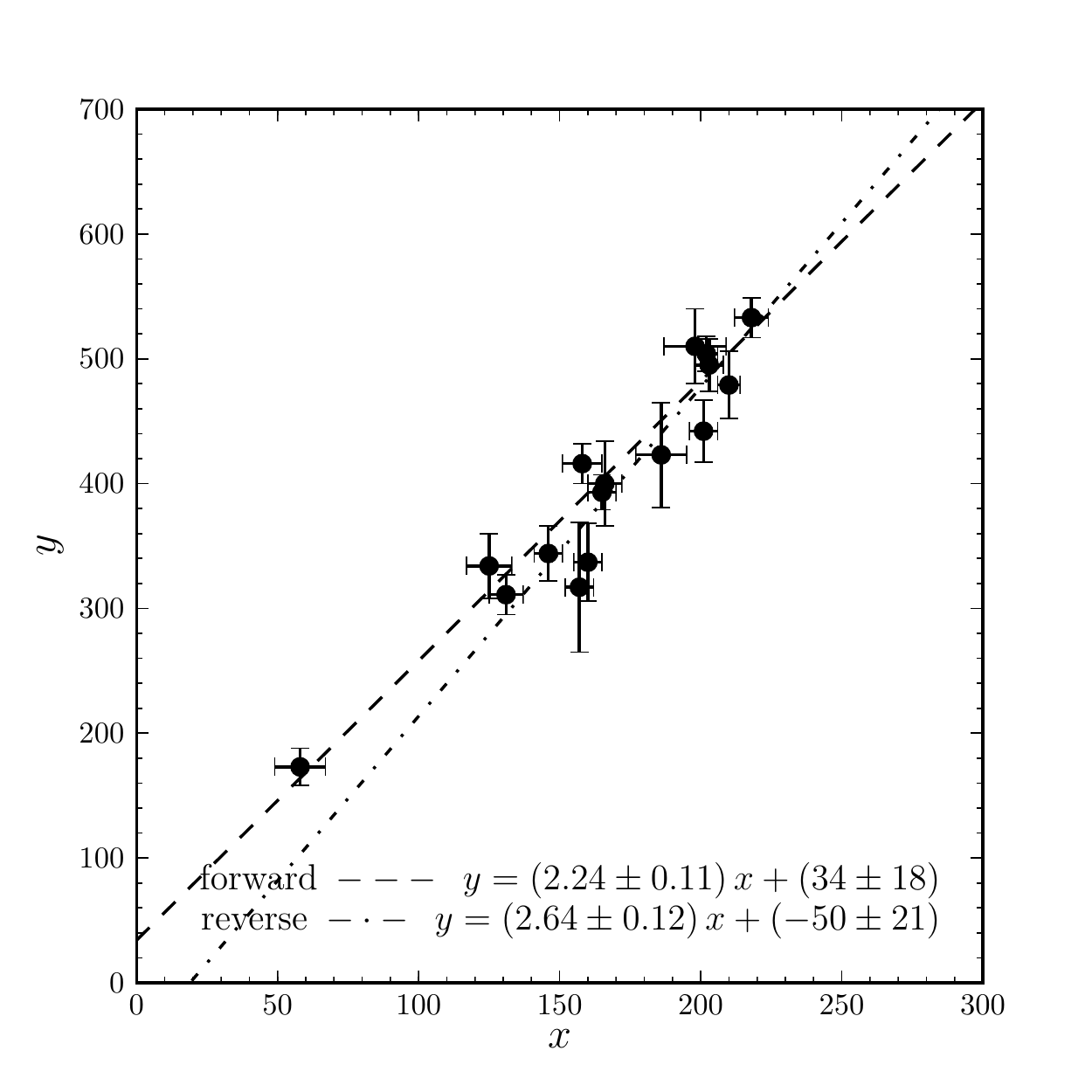}
\caption{Partial solution to \problemname~\ref{prob:forwardreverse}:
  Results of ``forward and reverse'' fitting.  Don't ever do
  this.}\label{fig:forwardreverse}
\end{figure}

\begin{problem}\label{prob:forwardreverse}
Perform the abominable forward--reverse fitting procedure on points 5
through 20 from \tablename~\ref{table:data_allerr} on
page~\pageref{table:data_allerr}.  That is, fit a straight line to the
$y$ values of the points, using the $y$-direction uncertainties
$\sigma_y^2$ only, by the standard method described in
\sectionname~\ref{sec:standard}.  Now transpose the problem and fit
the same data but fitting the $x$ values using the $x$-direction
uncertainties $\sigma_x^2$ only.  Make a plot showing the data points,
the $x$-direction and $y$-direction uncertainties, and the two
best-fit lines.  Your plot should look like
\figurename~\ref{fig:forwardreverse}.  Comment.
\end{problem}

\begin{figure}[htbp]
\exampleplot{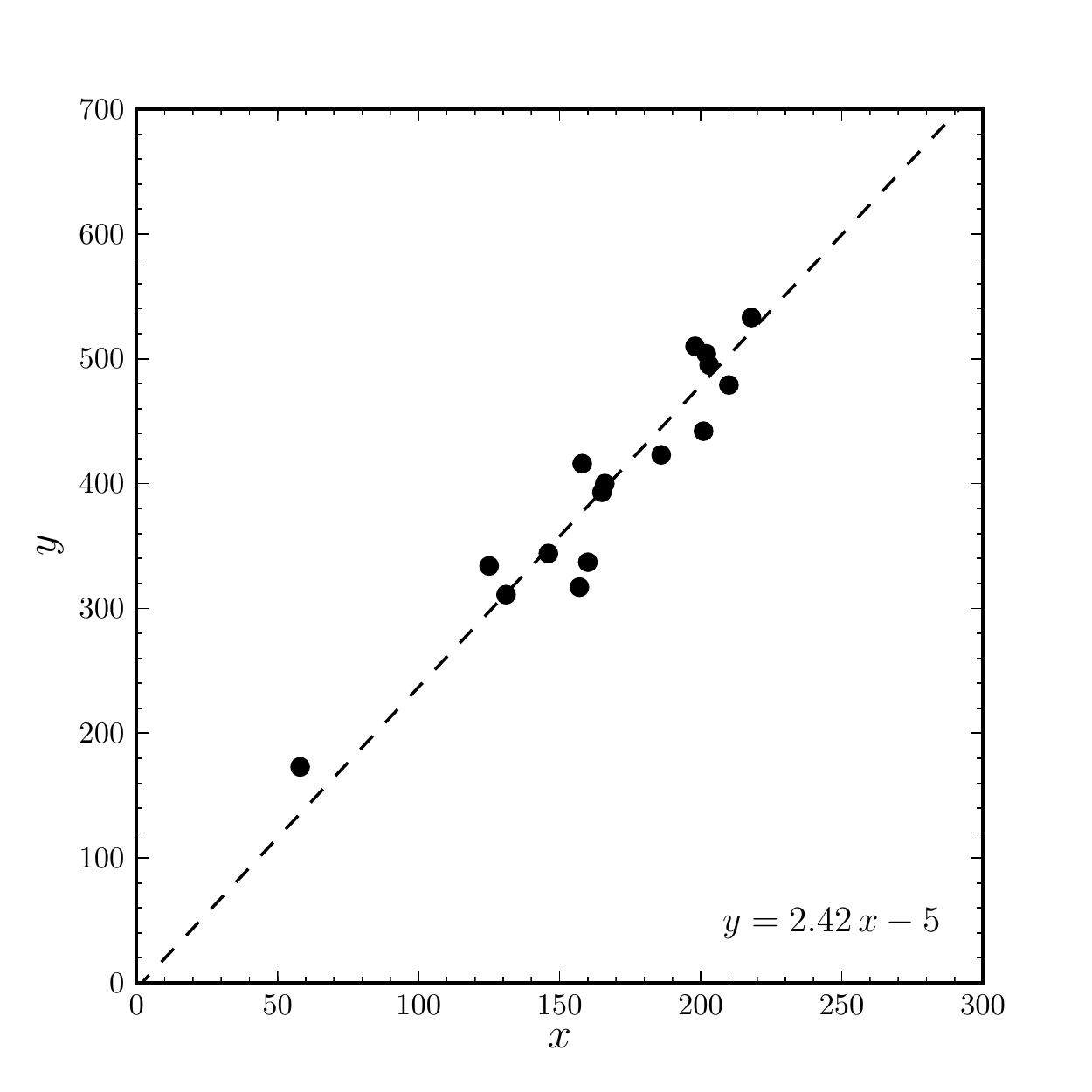}
\caption{Partial solution to \problemname~\ref{prob:pca}: The dominant
component from a principal components analysis.}\label{fig:pca}
\end{figure}

\begin{problem}\label{prob:pca}
Perform principal components analysis on points 5 through 20 from
\tablename~\ref{table:data_allerr} on
page~\pageref{table:data_allerr}.  That is, diagonalize the $2\times
2$ matrix $\mQ$ given by
\begin{equation}
\mQ = \sum_{i=1}^N\,\left[\mZ_i-\meanZ\right]
  \,\transpose{\left[\mZ_i-\meanZ\right]} \quad ,
\end{equation}
\begin{equation}
\meanZ = \frac{1}{N}\,\sum_{i=1}^N\,\mZ_i
\end{equation}
Find the eigenvector of $\mQ$ with the largest eigenvalue.  Now make a
plot showing the data points, and the line that goes through the mean
$\meanZ$ of the data with the slope corresponding to the direction of
the principal eigenvector.  Your plot should look like
\figurename~\ref{fig:pca}.  Comment.
\end{problem}

\section{Intrinsic scatter}\label{sec:scatter}

So far, everything we have done has implicitly assumed that there
truly is a narrow linear relationship between $x$ and $y$ (or there
would be if they were both measured with negligible uncertainties).
The words ``narrow'' and ``linear'' are both problematic, since there
are very few problems in science, especially in astrophysics, in which
a relationship between two observables is expected to be either.  The
reasons for intrinsic scatter abound; but generically the quantities
being measured are produced or affected by a large number of
additional, unmeasured or unmeasurable quantities; the relation
between $x$ and $y$ is rarely exact, even in when observed by
theoretically perfect, noise-free observations.

Proper treatment of these problems gets into the complicated area of
estimating density given finite and noisy samples; again this is a
huge subject so we will only consider one simple solution to the one
simple problem of a relationship that is linear but not narrow.  We
will not consider anything that looks like subtraction of variances in
quadrature; that is pure procedure and almost never
justifiable.\note{The standard method for estimating the width of a
  linear relationship is to compare the scatter of the data points to
  the scatter expected if the relationship is narrow.  The estimated
  intrinsic scatter can be ``found'' in this case by a procedure of
  subtracting in quadrature (intrinsic variance is observed variance
  minus expected variance given observational uncertainties).  This is
  not a good idea in general.  It doesn't work at all if the data
  points have a significant dynamic range in their measurement
  uncertainties, it can produce an estimate of the intrinsic scatter
  variance that is negative, and it returns no confidence interval or
  uncertainty.  In the case of negligible $x$-direction uncertainties,
  a better method is to add a parameter $V_y$ which is a variance to
  be added in quadrature to every data-point uncertainty variance;
  this is what we advocate in the main text.}  Instead---as usual---we
introduce a model that generates the data and infer the parameters of
that model.

Introduce an intrinsic Gaussian variance $V$, orthogonal to the line.
In this case, the parameters of the relationship become
$(\theta,\bperp,V)$.  In this case, each data point can be treated as
being drawn from a projected distribution function that is a
\emph{convolution} of the projected uncertainty Gaussian, of variance
$\Sigma_i^2$ defined in \equationname~(\ref{eq:Sigma}), with the
intrinsic scatter Gaussian of variance $V$.  Convolution of Gaussians
is trivial and the likelihood in this case becomes
\begin{equation}
\ln\like = K - \sum_{i=1}^N \frac{1}{2}\,\ln(\Sigma_{i}^2+V)
 - \sum_{i=1}^N \frac{\Delta_i^2}{2\,[\Sigma_{i}^2+V]} \quad ,
\end{equation}
where again $K$ is a constant, everything else is defined as it is in
\equationname~(\ref{eq:twodlike}), and an additional term has
appeared to penalize very broad variances (which, because they are
less specific than small variances, have lower likelihoods).
Actually, that term existed in \equationname~(\ref{eq:twodlike}) as
well, but because there was no $V$ parameter, it did not enter into
any optimization so we absorbed it into $K$.

As we mentioned in \notename~\ref{note:orthogonal}, there are
limitations to this procedure because it models only the distribution
\emph{orthogonal} to the relationship.\note{One confusing issue in the
  subject of intrinsic scatter is the \emph{direction} for the
  scatter.  If things are Gaussian, it doesn't matter whether you
  think of the scatter being in the $x$ direction, the $y$ direction,
  or the direction perpendicular to the line (as we think of it in the
  method involving $V$ in this \sectionname).  However it must be made
  clear which direction is being used for the statement of the result,
  and what the conversions are to the other directions (these involve
  the angle $\theta$ or slope $m$ of course).}  Probably all good
methods fully model the intrinsic two-dimensional density function,
the function that, when convolved with each data point's intrinsic
uncertainty Gaussian, creates a distribution function from which that
data point is a single sample.  Such methods fall into the
intersection of density estimation and deconvolution, and completely
general methods exist.\note{Density estimation by generative modeling,
  in situations of heteroscedastic data, has been solved for
  situations of varying generality by \cite{kelly07} and \cite{bovy}.}

\begin{figure}[htbp]
\exampleplot{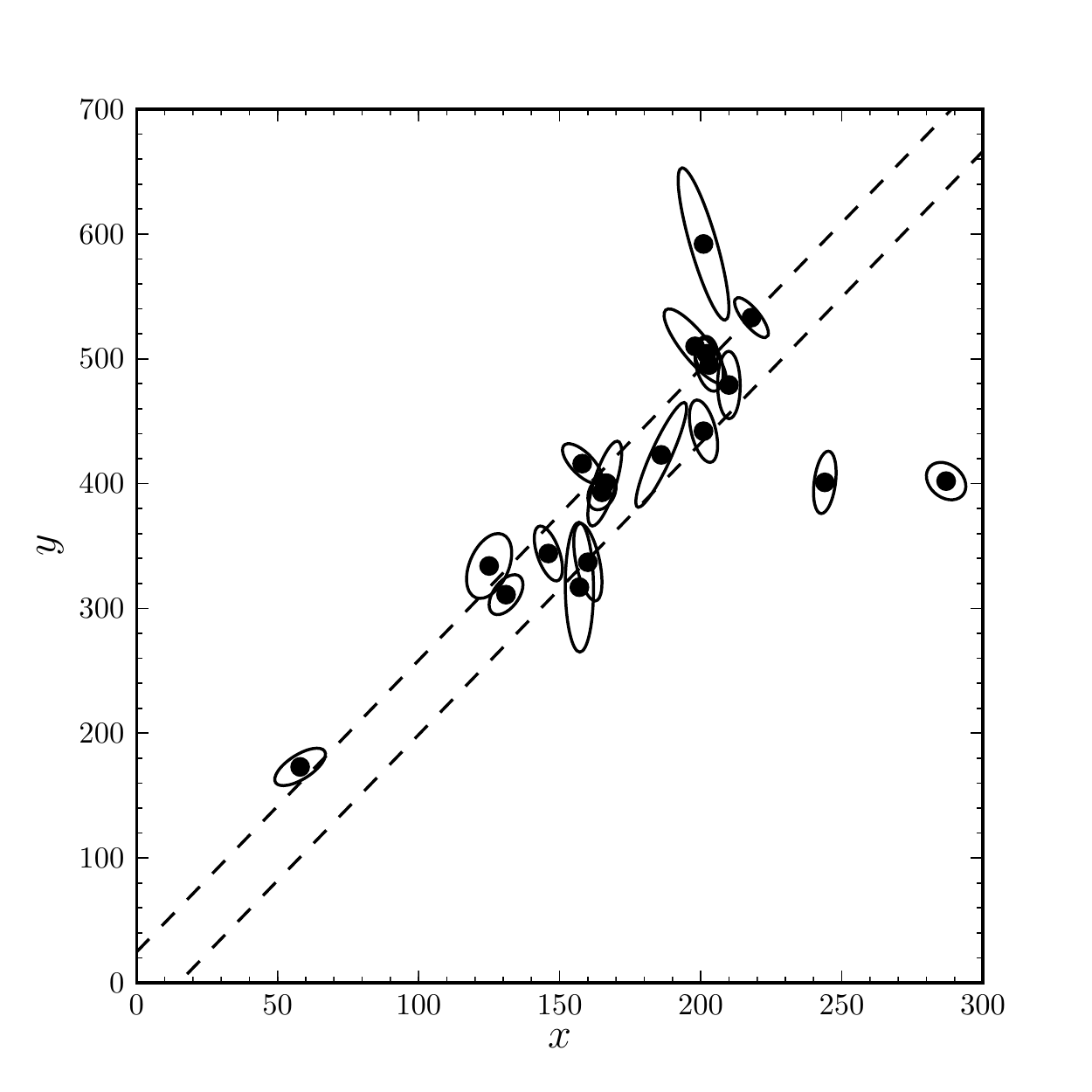}
\caption{Partial solution to \problemname~\ref{prob:intrinsic}: The
maximum-likelihood fit, allowing for intrinsic
scatter.}\label{fig:intrinsic}
\end{figure}

\begin{problem}\label{prob:intrinsic}
Re-do \problemname~\ref{prob:twod}, but now allowing for an orthogonal
intrinsic Gaussian variance $V$ and only excluding data point 3.
Re-make the plot, showing not the best-fit line but rather the
$\pm\sqrt{V}$ lines for the maximum-likelihood \emph{intrinsic}
relation.  Your plot should look like \figurename~\ref{fig:intrinsic}.
\end{problem}

\begin{figure}[htbp]
\exampleplot{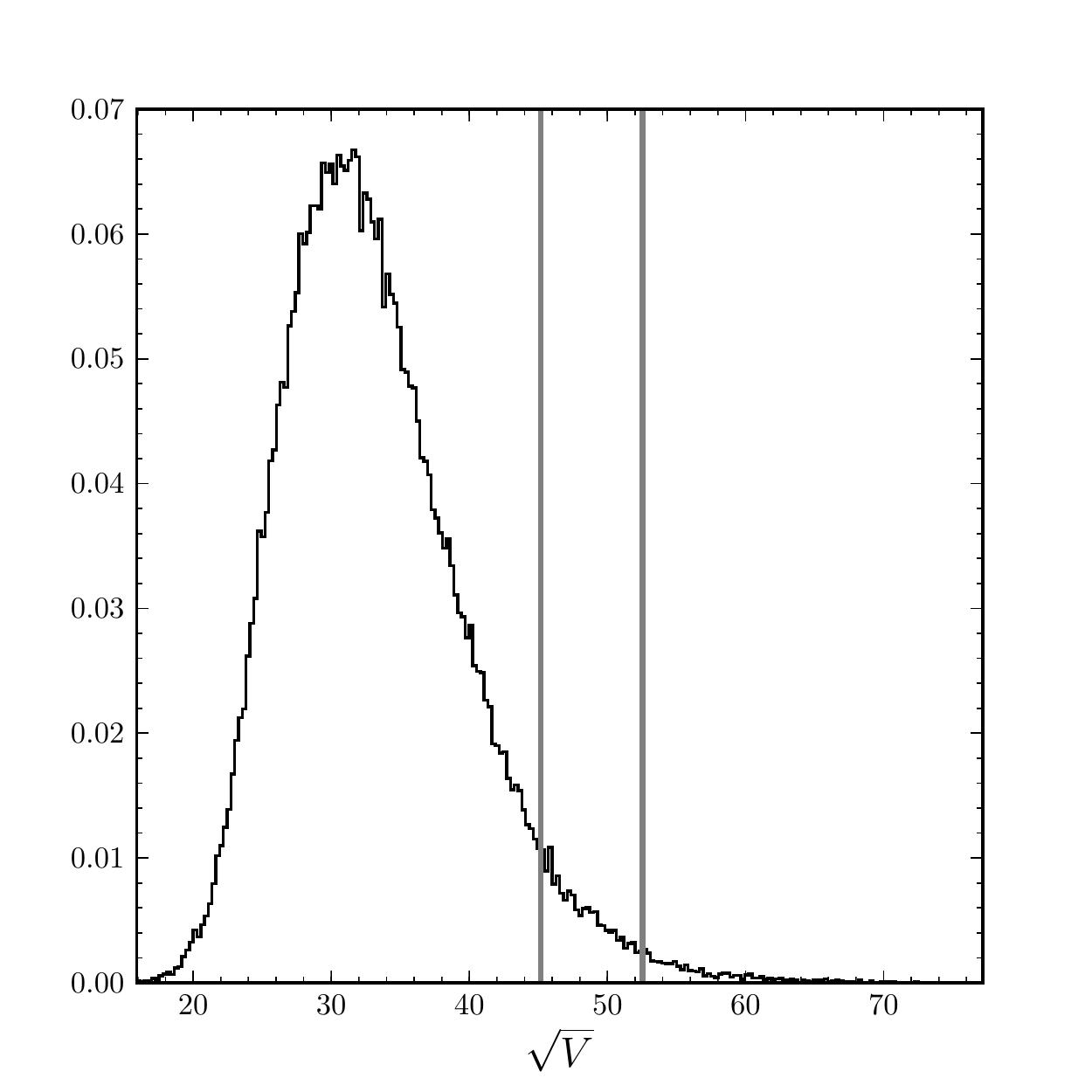}
\caption{Partial solution to \problemname~\ref{prob:bayesintrinsic}:
The marginalized posterior probability distribution for the intrinsic
variance.}\label{fig:bayesintrinsic}
\end{figure}

\begin{problem}\label{prob:bayesintrinsic}
Re-do \problemname~\ref{prob:intrinsic} but as a Bayesian, with
sensible Bayesian priors on $(\theta,\bperp,V)$.  Find and marginalize
the posterior distribution over $(\theta,\bperp)$ to generate a
marginalized posterior probability for the intrinsic variance
parameter $V$.  Plot this posterior with the 95 and 99~percent
\emph{upper limits} on $V$ marked.  Your plot should look like
\figurename~\ref{fig:bayesintrinsic}.  Why did we ask only for upper
limits?
\end{problem}

\clearpage
\markright{Notes}\theendnotes


\clearpage
\begin{thebibliography}{}\markright{References}
\bibitem[Bovy, Hogg, \& Roweis(2009)]{bovy}
  Bovy,~J., Hogg,~D.~W., \& Roweis, S.~T., 2009,
  Extreme deconvolution: inferring complete distribution functions from noisy, heterogeneous, and incomplete observations, 
  arXiv:0905.2979 [stat.ME]
\bibitem[Jaynes(2003)]{jaynes}
  Jaynes,~E.~T., 2003,
  \textit{Probability theory: the logic of science} (Cambridge University Press)
\bibitem[Gilks, Richardson, \& Spiegelhalter(1995)]{gilksmcmc}
  Gilks,~W.~R., Richardson,~S., \& Spiegelhalter,~D., 1995,
  \textit{Markov chain Monte Carlo in practice: interdisciplinary statistics} (Chapman \& Hall/CRC)
\bibitem[Hampel \etal(1986)]{hampel}
  Hampel,~F.~R., Ronchetti,~E.~M., Rousseeuw,~P.~J., \& Stahel,~W.~A., 1986, 
  \textit{Robust statistics: the approach based on influence functions} (New York: Wiley)
\bibitem[Isobe \etal(1990)]{isobe90}
  Isobe,~T., Feigelson, E.~D., Akritas,~M.~G., \& Babu,~G.~J., 1990,
  Linear regression in astronomy,
  \textit{Astrophysical Journal} \textbf{364} 104
\bibitem[Kelly(2007)]{kelly07}
  Kelly,~B.~C., 2007,
  Some aspects of measurement error in linear regression of astronomical data,
  \textit{Astrophysical Journal} \textbf{665} 1489
\bibitem[Mackay(2003)]{mackay}
  Mackay,~D.~J.~C., 2003,
  \textit{Information theory, inference, and learning algorithms} (Cambridge University Press)
\bibitem[Neal(2003)]{neal2003a}
  Neal.,~R.~M., 2003,
  Slice sampling,
  \textit{Annals of Statistics}, \textbf{31}(3), 705
\bibitem[Press(1997)]{pressH0}
  Press,~W.~H., 1997,
  Understanding data better with Bayesian and global statistical methods,
  in \textit{Unsolved problems in astrophysics,}
  eds. Bahcall,~J.~N. \& Ostriker,~J.~P. (Princeton University Press)
  49--60
\bibitem[Press \etal(2007)]{press}
  Press,~W.~H., Teukolsky,~S.~A., Vetterling,~W.~T., \& Flannery,~B.~P., 2007,
  \textit{Numerical recipes: the art of scientific computing} (Cambridge University Press)
\bibitem[Sivia \& Skilling(2006)]{sivia}
  Sivia,~D.~S. \& Skilling,~J., 2006,
  \textit{Data analysis: a Bayesian tutorial} (Oxford University Press)
\end{thebibliography}
\end{document}